\documentclass[aps,prx,twocolumn,superscriptaddress,notitlepage,longbibliography]{revtex4-1}
\usepackage{subdepth}
\usepackage{xcolor}

\usepackage{graphicx}
\usepackage{epsfig}
\usepackage{epstopdf}
\usepackage{float}
\usepackage{subfigure}
\usepackage{bbold}

\usepackage{dcolumn}
\usepackage{latexsym}
\usepackage{amsmath,amsfonts,amsthm,bm}
\usepackage{wasysym}
\usepackage{bm}

\usepackage[colorlinks,bookmarks=false,citecolor=blue,linkcolor=red,urlcolor=blue]{hyperref}
\usepackage{verbatim}

\usepackage{booktabs}

\usepackage{ulem} 

\usepackage{algorithm}
\usepackage{algorithmic}
\usepackage{indentfirst}
\usepackage{listings}

\newcommand{\RNum}[1]{\uppercase\expandafter{\romannumeral #1\relax}}

\def\be{\begin{equation}}
\def\ee{\end{equation}}
\def\bea{\begin{eqnarray}}
\def\eea{\end{eqnarray}}

\usepackage{array}
\newcommand{\PreserveBackslash}[1]{\let\temp=\\#1\let\\=\temp}
\newcolumntype{C}[1]{>{\PreserveBackslash\centering}p{#1}}
\newcolumntype{R}[1]{>{\PreserveBackslash\raggedleft}p{#1}}
\newcolumntype{L}[1]{>{\PreserveBackslash\raggedright}p{#1}}

\usepackage{color}

\definecolor{darkblue}{rgb}{0,0.02,0.45}
\definecolor{darkred}{rgb}{0.45,0.02,0} 

\makeatletter
\newcommand{\thickhline}{%
\noalign {\ifnum 0=`}\fi \hrule height 0.7pt
\futurelet \reserved@a \@xhline
}
\newcolumntype{"}{@{\hskip\tabcolsep\vrrule width 0.7pt\hskip\tabcolsep}}
\makeatother

\usepackage{times}

\begin{document}
\title{Phonon Dynamics in the  Chiral Kitaev Spin Liquid}
\author{Susmita Singh}
\affiliation{School of Physics and Astronomy, University of Minnesota, Minneapolis, MN 55455, USA}
\author{P. Peter Stavropoulos}
\affiliation{School of Physics and Astronomy, University of Minnesota, Minneapolis, MN 55455, USA}
\author{Natalia B. Perkins}
\affiliation{School of Physics and Astronomy, University of Minnesota, Minneapolis, MN 55455, USA}

\date{\today}

\begin{abstract}


We investigate the effect of a magnetic field on the extended Kitaev spin liquid state through phonon dynamics. Using a constrained fermionic self-consistent mean field method, we analyze the quantum spin liquid (QSL) ground state for the extended Kitaev model with both the Zeeman term and the perturbative three-spin interaction term $\kappa$. Our results demonstrate the dependence of the stability of the Kitaev QSL state on the field direction, consistent with findings in the literature. Additionally, we calculate the phonon dynamics for acoustic phonons coupled to the Majorana fermion excitations of the Kitaev spin liquid state, discussing the temperature and field evolution of these quantities.

\end{abstract}

\maketitle
\section{Introduction}

 The Kitaev model on the honeycomb lattice \cite{Kitaev2006} has played a central role in the studies  of 
  QSLs,
 serving as one of the simplest yet exactly solvable effective models. This model not only realizes the QSL state but is also pertinent to real materials \cite{Jackeli2009,Chaloupka2010,Knolle2017,Takagi2019,Trebst2022,Rousochatzakis2024}.
The model arises in tricoordinated geometries with magnetic ion sites that are well described by pseudo-spin $J_{\text{eff}} = 1/2$ Kramers' doublets \cite{Jackeli2009}. These conditions are most closely realized  in the layered compounds A$_2$IrO$_3$ (A=Na, Li)~\cite{Yogesh2010,Liu2011,Choi2012,
Ye2012,Williams2016} and $\alpha$-RuCl$_3$~\cite{Plumb2014,Sears2015,Majumder2015,Banerjee2016}. In these systems,  
magnetic ions predominantly interact via nearest-neighbor (NN) Ising-like interactions, with quantization axes that depend on the  
 bond's direction,
mimicking the Kitaev interactions  (hence the term ‘Kitaev materials’).  However, most of these Kitaev materials exhibit magnetic ordering at sufficiently low temperatures, indicating the presence of additional interactions and the fragility  of the  Kitaev QSL against various perturbations.

 Recent  experimental  efforts  have been focused on applying a magnetic field to these systems.  In particular interesting results were obtained in experiments on the Kitaev  candidate material $\alpha$-RuCl$_3$, in which an in-plane  magnetic field destroys the zero-field zigzag  magnetic ordering and
 reveals 
 a transition into a quantum paramagnetic phase with a spin-excitation gap
\cite{Ponomaryov2017,baek_evidence_2017,banerjee_excitations_2018,BalzPRB2019,gass_field-induced_2020,Sahasrabudhe2020}.
  These observations triggered a flurry of theoretical work to comprehend the behavior of Kitaev materials in a magnetic field \cite{Jiang2011PRB,Janssen2016,Chern2017,Liang2018,Gohlke2018,Hickey2019,Janssen2019,Ralko2020,Shangshun2021,Yifm2022,Li2022,Franke2022,Joy2022,Saheli2023}. 
  However, despite considerable effort, the precise nature of this field-induced disordered magnetic phase is not yet known. Several experimental studies \cite{baek_evidence_2017,banerjee_excitations_2018}, and particularly the controversial observation of an approximately half-integer quantized thermal Hall effect in $\alpha$-RuCl$_3$ at around 8T  \cite{kasahara_majorana_2018,Yamashita2020,Yokoi2021,bruin2022}, suggested that this phase is a chiral QSL coupled to phonons \cite{Ye2018,Aviv2018}. 
  
  Motivated by these findings, the potential of phonon dynamics as a probe to study fractionalized excitations in the Kitaev spin liquid was explored both experimentally in $\alpha$-$\text{RuCl}_3$ \cite{LebertPRB2022,Miao2020,Hauspurg2024} and theoretically in generic Kitaev spin models \cite{Mengxing2020,Metavitsiadis2020,Kexin2021-s,Kexin2021,Metavitsiadis2022,Feng2022, 
  Kocsis2022,Susmita2023}. 
  In  particular, in our previous study \cite{Susmita2023}, we examined how the features of fractionalization can be observed in the sound attenuation in the generalized $J$-$K$-$\Gamma$ model \cite{Rau2014,Sizyuk2014,Wang2017,Ran2017}.
  This model,  in addition
to the Kitaev  interaction ($K$), includes
   the Heisenberg interaction ($J$), and the symmetric off-diagonal interaction ($\Gamma$). While not always sufficient, the $J$-$K$-$\Gamma$ model is often regarded as the minimal model describing the Kitaev candidate materials \cite{Winter2017,MaksimovPRR2020,Rousochatzakis2024}. This extended Kitaev model is no longer exactly solvable, but by employing various numerical and perturbative methods \cite{Schaffer2012,Knolle2018,Gohlke2018,Gohlke2018,Ralko2020,Shangshun2021,Joy2022,Cookmeyer2023,Susmita2023},   it was shown 
 that while the  Kitaev QSL persists for small $J$ and $\Gamma$ interactions, the transition to  magnetically ordered phases occurs when these terms become strong enough.

In this paper, we investigate the response of the extended Kitaev spin liquid  to a magnetic field through the dynamics of acoustic phonons. Apart from breaking time reversal symmetry (TRS), the effects of the magnetic field $h$ on phonon dynamics originate from two factors: to linear order in $h$, the Zeeman coupling induces a hopping of visons, making them dynamic variables, and to cubic order in $h$, the so-called $\kappa$-interaction term induces a gap in the Majorana spectrum. It was shown that the spectral gap can also be opened by the $\Gamma'$ term \cite{Takikawa2019}, but we constrain our analysis to the effects of gap opening by the magnetic field only. Given that the addition of the Zeeman field even to the pure Kitaev model breaks its exact solvability, in this study, we explore phonon dynamics using the 
constrained
self-consistent Majorana mean-field theory \cite{Knolle2018,Ralko2020,Joy2022,Susmita2023}.
We compute two observables related to phonon dynamics -- the sound attenuation ($\alpha_s$) and the Hall viscosity ($\eta_H$). Although both these quantities can be connected to the phonon polarization bubble \cite{Mengxing2020}, 
 their origins are different. 
Sound attenuation originates from dissipative scattering processes,  while the Hall viscosity emerges from non-dissipative mechanism and requires TRS breaking. 
 We discuss in detail 
the dependence of $\alpha_s$ and $\eta_H$ on $h$ and $\kappa$, as well as their temperature evolution, and  show that these two quantities 
 exhibit distinct characteristic behaviors.
Finally, motivated  by the sound attenuation experiments 
in $\alpha$-RuCl$_3$ under  in-plane magnetic field  along the ${\bf a}$ crystallographic axis \cite{Hauspurg2024},
we explore the phonon dynamics of the generalized $J$-$K$-$\Gamma$ model in this specific geometry.

\section{Model}
In this section, we describe the components of our system, which comprises a quantum spin liquid interacting with lattice vibrations (acoustic phonons) in the presence of an external magnetic field.
The spin Hamiltonian is reformulated in terms of Majorana fermions, leading to a mean-field description that captures the spin liquid state within a finite parameter region of the extended Kitaev model. 
We then couple the spins and acoustic phonons by writing the 
symmetry-allowed interactions between them, which are subsequently recast into Majorana language.

\subsection{Spin Hamiltonian}
\begin{figure}
	\centering
	\includegraphics[width=0.98\columnwidth]{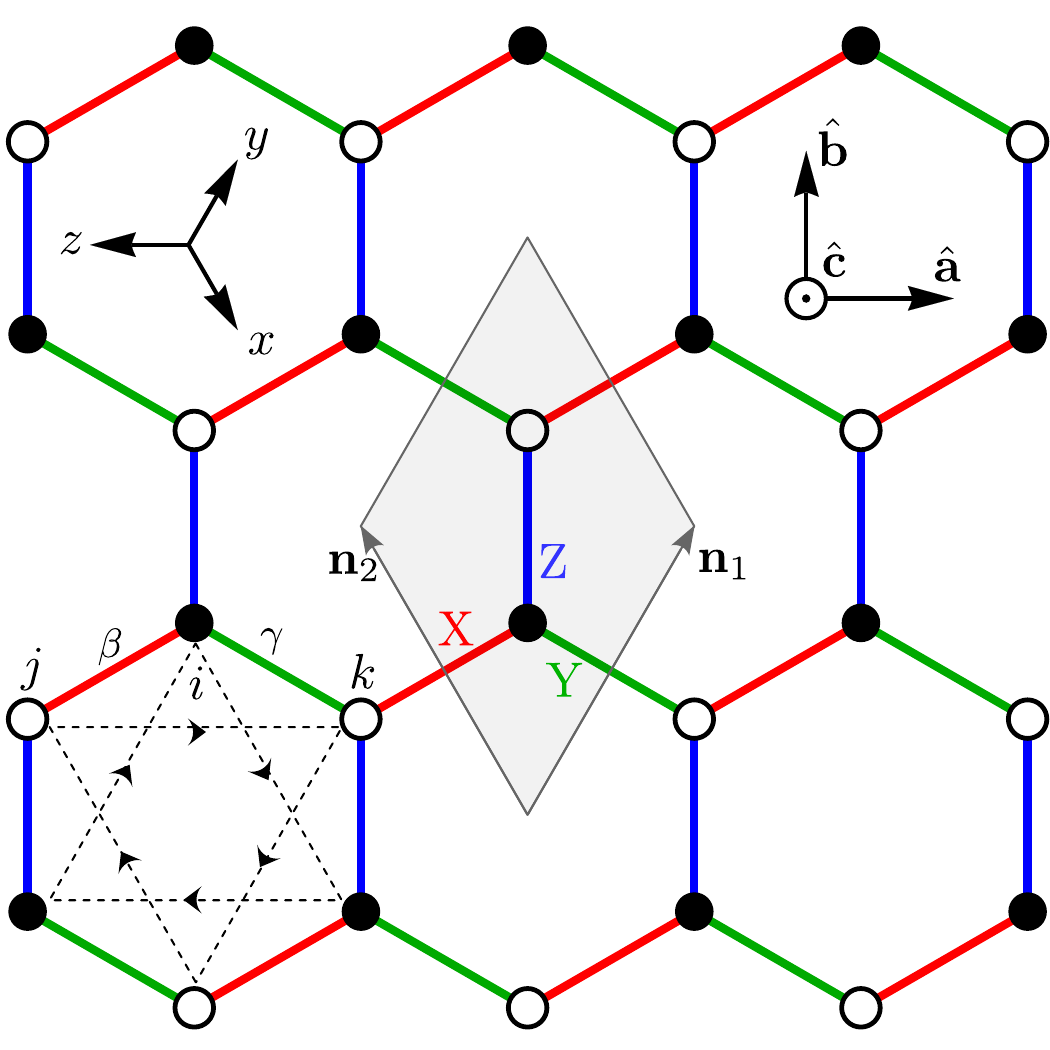}
     \caption{The honeycomb unit cell defined by lattice vectors $\mathbf{n}_1$ and $\mathbf{n}_2$ with two sublattices, $A$ in white and $B$ in black. At the top left the cubic axis $xyz$ are shown, along with the crystalographic axis on the top right, with $\hat{\mathbf{a}}=[1,1,-2]/\sqrt{6}$ and $\hat{\mathbf{b}}=[-1,1,0]/\sqrt{2}$ defining the honeycomb plane, and the out of plane direction $\hat{\mathbf{c}}=[1,1,1]/\sqrt{3}$. The mean field decoupled $\kappa$ term involves NNN hopping, shown on the bottom left, and one example of index setup shown.}
\label{Fig:basic_honey_figure}
\end{figure}

We start our discussion by considering the  extended  Kitaev model on the honeycomb lattice in the presence of a uniform magnetic field. The
 Hamiltonian  has three contributions:
\begin{equation}\label{eq:spin_ham}
    \mathcal{H}_s \! =\! \mathcal{H}_0+\mathcal{H}_{\mathcal{Z}}+\mathcal{H}_{\kappa},
\end{equation}   
with   
 \begin{equation}\label{eq:spin_ham-terms}
\begin{array}{c}   
    \multicolumn{1}{l}{\mathcal{H}_0\!=\!\displaystyle\sum_{\langle ij \rangle_{\alpha}} 
    \Big[ J \bm{\sigma}_{i}\!\cdot\!\bm{\sigma}_{j} \!\!+\! K\sigma^{\alpha}_{i}\sigma^{\alpha}_{j} + \Gamma\left(\sigma^{\beta}_{i}\sigma^{\gamma}_{j} \!\!+\!\sigma^{\gamma}_{i}\sigma^{\beta}_{j}\right)\!\! \Big]},\\[12pt]
    \multicolumn{1}{l}{\mathcal{H}_{\mathcal{Z}} \!=\! -\displaystyle\sum_{i\in A,B} \mathbf{h}\cdot\bm{\sigma}_{i}\,,}\\[12pt]
    \multicolumn{1}{l}{\mathcal{H}_{\kappa} \!=\!  -\displaystyle\sum_{\langle ijk \rangle}\kappa\sigma_{i}^{\alpha}\sigma_{j}^{\beta}\sigma_{k}^{\gamma}\,,}
\end{array}
\end{equation}
where $\mathcal{H}_0$ is  the $J$-$K$-$\Gamma$ model, $\mathcal{H}_{\mathcal{Z}}$ is the Zeeman  coupling with $\mathbf{h}=h \hat{\mathbf{h}}$ denoting  magnetic field  in an arbitrary  direction $\hat{\mathbf{h}}$, 
 and $\mathcal{H}_{\kappa}$ 
 is the three-spin-interaction term derived pertubatively 
by considering 
$h\ll K$ \cite{Kitaev2006}, with $\kappa\sim \frac{h_xh_yh_z}{K^2}$ appearing in the third order in $h$. Here we used the following notations:
 $\bm{\sigma}=(\sigma^x,\sigma^y,\sigma^z)$ are the Pauli matrices, $\langle ij \rangle_\alpha$ are the nearest-neighbor (NN) bonds with  $\alpha=x,y,z$ and $\langle ijk \rangle$ is defined between three sites as shown in Fig.~\ref{Fig:basic_honey_figure}.

\subsubsection{Symmetry} \label{sec:symmetry}
 Let us discuss the symmetries of the   Hamiltonian Eq.~(\ref{eq:spin_ham}).
  In the absence of the magnetic field, 
the $J$-$K$-$\Gamma$ Hamiltonian $\mathcal{H}_0$ is time reversal  invariant and respects the crystallographic point group $D_{3d}$. This point group comprises of $C_3$ rotations around the $\hat{\mathbf{c}}\propto[1,1,1]$ axis, $C_{2{\text{X(Y,Z)}}}$ rotations around the $\text{X(Y,Z)}$-bond direction, inversion $I$, and mirror planes $m_{{\text{X(Y,Z)}}}=IC_{2{\text{X(Y,Z)}}}$.
  The presence of the magnetic field not only breaks TRS ($\mathcal{T}$),  but also
  breaks the point group symmetry for a generic direction of the magnetic field, 
   except when $\mathbf{h} \parallel  \hat{\mathbf{c}}$,
  for which the symmetry group remains isomorphic to $D_{3d}$.
   In this latter case,  
$C_3$ symmetry is preserved, as the rotation is performed along the field direction, and while  $C_{2{\text{X(Y,Z)}}}$ symmetries are broken, the combined operations $\mathcal{T}C_{2{\text{X(Y,Z)}}}$ remain a valid symmetry operation.
The key reason for this is that the $C_{2{\text{X(Y,Z)}}}$ rotations  have rotation axes perpendicular to the magnetic field direction $[1,1,1]$, and the additional time reversal operation flips the spins and  effectively restores the system to its original configuration. 
When the magnetic field $\mathbf{h}||\hat{\mathbf{a}}\propto[1,1,-2]$, the symmetry of the system is reduced to a group isomorphic to $C_{2h}$:  $C_3$, $C_{2X}$  and  $C_{2Y}$ rotations are no longer symmetry operations, but $\mathcal{T}C_{2\text{Z}}$ remains a symmetry.

\subsubsection{Mean-field decoupling}

In the presence of additional non-Kitaev interactions and the Zeeman field, the spin Hamiltonian  Eq.~(\ref{eq:spin_ham}) is no longer exactly solvable. In our analysis, we adopt a mean-field parton description, which is conveniently formulated using the fermionic parton decomposition. Building upon the original work by Kitaev \cite{Kitaev2006}, we represent the spin using four Majorana fermions 
\begin{equation}\label{eq:spin_to_maj}
\sigma^{\alpha} = i c^{\alpha}c^{o},\,
\{c^{a},c^{b}\}=2\delta_{a,b},\,(c^{a})^{\dagger}=c^{a},
\end{equation}
 with Greek indices $\alpha,\beta,\gamma$ running over $x,y,z$ and  Latin indices $a,b$ over $o,x,y,z$. Because of the enlarged Hilbert space, it becomes necessary to impose specific constraints to ensure that we confine our analysis within the physical subspace.  These constraints are
\begin{equation}\label{eq:constrains}
    \begin{array}{cc}
    i c^{\alpha} c^o \!+\! \dfrac{1}{2} \varepsilon^{\alpha\beta\gamma} i c^{\beta} c^{\gamma} \!=\! 0.
    \end{array}
\end{equation}
By pairing up Majoranas into complex fermions, 
the above constraint can be easily proven  by projecting out the vacuum and the double occupancy states (see details  in App.~\ref{appx:peter_you_better_get_this_appendix_done_already}).

 In terms of  Majorana fermions, the   spin Hamiltonian 
 contains  both bilinear and quartic terms. 
 The quartic terms appearing  in $\mathcal{H}_0$ and $\mathcal{H}_\kappa$
 can be effectively decoupled  by introducing a set of 
 mean fields
 and applying
 the Hubbard-Stratonovich transformation \cite{Hubbard1959,Stratonovich957}. The  mean-field decoupling of  $\mathcal{H}_0$  is obtained as follows 
\begin{equation}\label{eq:mf_decouple}
    \begin{array}{l}
        \sigma^{\alpha}_{i}\sigma^{\beta}_{j} = (i c^{\alpha}_{i}c^{o}_{i})(i c^{\beta}_{j}c^{o}_{j}) \xrightarrow[\text{decouple}]{\text{MF}} \\[0.5cm]
        \multicolumn{1}{r}{\left(2 i c^{\alpha}_{i}c^{o}_{i}\mathrm{m}^{\beta}_B + 2i c^{\beta}_{j}c^{o}_{j} \mathrm{m}^{\alpha}_A  - 4 \mathrm{m}^{\alpha}_A\mathrm{m}^{\beta}_B\right)} \\[0.3cm]
        \multicolumn{1}{r}{-\left( i c^{\alpha}_{i}c^{\beta}_{j}\mathrm{\Phi}^{oo}_{\delta} + i c^{o}_{i}c^{o}_{j}\mathrm{\Phi}^{\alpha\beta}_{\delta}  -  \mathrm{\Phi}^{oo}_{\delta}\mathrm{\Phi}^{\alpha\beta}_{\delta}\right)} \\[0.3cm]
        \multicolumn{1}{r}{+\left( i c^{\alpha}_{i}c^{o}_{j}\mathrm{\Phi}^{o\beta}_{\delta} + i c^{o}_{i}c^{\beta}_{j}\mathrm{\Phi}^{\alpha o}_{\delta}  -  \mathrm{\Phi}^{o\beta}_{\delta}\mathrm{\Phi}^{\alpha o}_{\delta}\right)},
    \end{array}
\end{equation}
where $\bm{\delta}=\bm{R}_j-\bm{R}_i$ is 
the lattice vector difference corresponding to X-, Y-, or Z- bond formed between sites  $i\in$ sublattice $A$  and $j\in$ sublattice $B$,  and the mean fields   are defined as
\begin{equation}\label{eq:mfv_in_rspace}
\begin{array}{c}
\mathrm{m}^{\alpha}_L \!=\! \dfrac{1}{2N}\!\displaystyle\sum_{i\in L}\!\langle ic^{\alpha}_{i}c^{o}_{i}  \rangle,\\[12pt]
\mathrm{\Phi}^{a b}_{\delta}\!=\!\dfrac{1}{N}\displaystyle\sum_{i\in A} \langle ic^{a}_{i}c^{b}_{i+\delta }  \rangle, \\[12pt]
\end{array}
\end{equation}
where $L$ denotes sublattice $A$ or $B$.
The factor of 1/2 in the magnetic channels $\mathrm{m}$ is introduced to normalize the magnetization vector $\mathbf{m}_L$ to a length of 1/2. 
 Note that the bond  mean-field parameter $\Phi_\delta$ is always defined from sublattice $A$ to sublattice $B$ on the $\delta$-bond.

 The decoupling procedure  in  $\mathcal{H}_{\kappa}$, involving
 three spins, reads 
\begin{equation}
\sigma_{i}^{\alpha}\sigma_{j}^{\beta}\sigma_{k}^{\gamma} \xrightarrow[]{} -ic^{o}_{j}c^{o}_{k} \Phi^{\beta\beta}_{\beta}\Phi^{\gamma\gamma}_{\gamma}
\end{equation}
resulting in an effective next nearest neighbor (NNN) hopping between  sites belonging to the same sublattice, with hopping strength determined by the NN mean-field values $\Phi_\beta$ and  $\Phi_\gamma$  on the bonds emanating from the shared site $i$.  

Collecting all the terms from above,  we obtain the resulting mean-field Hamiltonian 
in the  momentum space
\begin{equation}\label{eq:MF_spin_ham}
\begin{array}{c}
\mathcal{H}_s^{\text{MF}} \!=\! \displaystyle\sum_{\bm{k}} \mathcal{C}_{\bm{-k}}^{T} H^{\text{MF}}_{\bm{k}}\mathcal{C}_{\bm{k}},\\[12pt]
\mathcal{C}_{\bm{k}}^{T}\!=\! \left[c_{A,\bm{k}}^{a},  c_{B,\bm{k}}^{a} \right], 
\end{array}
\end{equation}
where the $8\times8$ 
matrix $H^{\text{MF}}_{\bm{k}}$ is a function of i) the Lagrange multipliers $\lambda_L^{\alpha}$, each  enforcing the fermionic constraints Eq.~(\ref{eq:constrains}) on average for each sublattice,  ii) sublattice magnetizations
$\mathbf{m}_L\!=\!(\mathrm{m}_L^{x},\mathrm{m}_L^{y}, \mathrm{m}_L^{z})$, 
iii) the NN bond fields $\mathrm{\Phi}^{a b}_{ \delta}$, and 
iv) model parameters $J,K,\Gamma,h,\kappa$. Further details on the derivation and the explicit form of $H^{\text{MF}}_{\bm{k}}$ can be found in App.~\ref{appx:peter_you_better_get_this_appendix_done_already}.

 \begin{center}
\begin{figure*}
	\centering
	\includegraphics[width=0.95\linewidth]{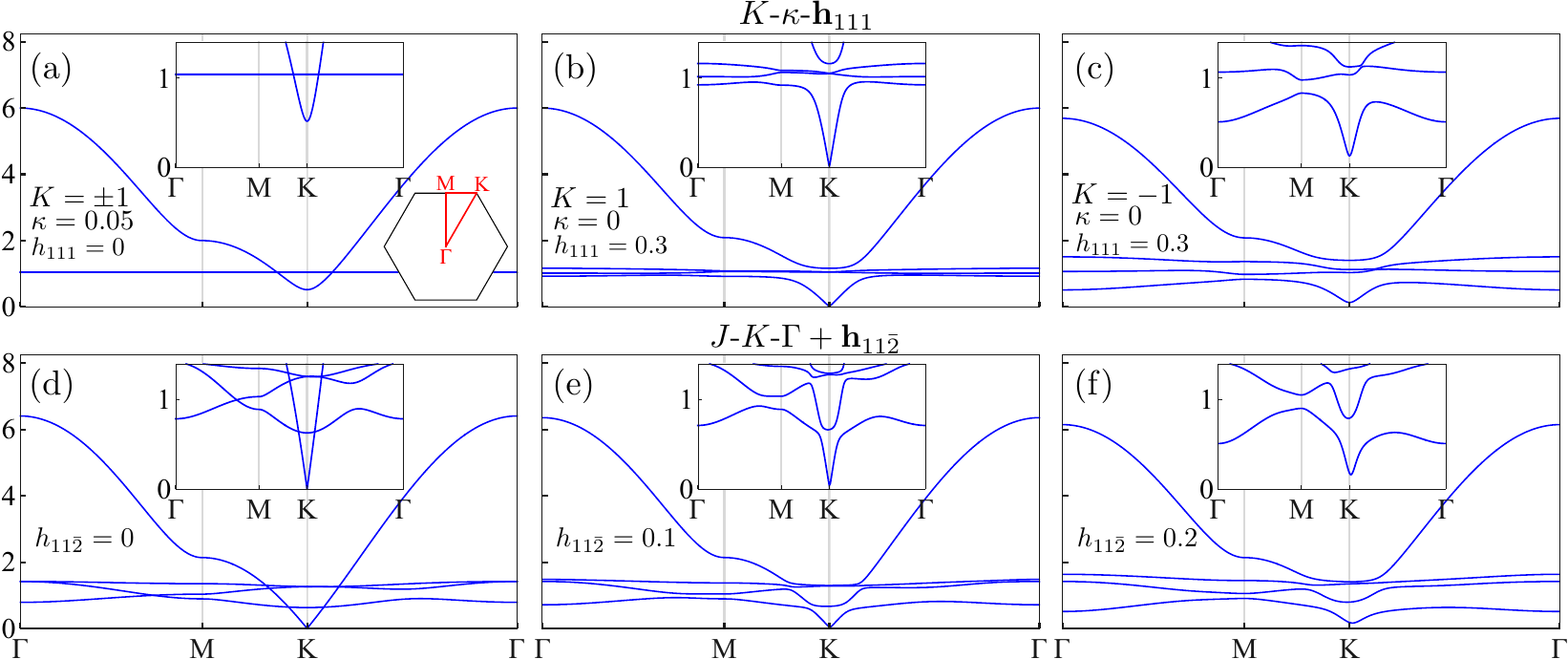}
     \caption{The mean-field spectrum  for the extended Kitaev model  in the presence of  magnetic field along the BZ path $\Gamma-M-K-\Gamma$, shown in inset of (a).  (a) The MF spectrum for the the $K+\kappa$ model agrees with the exact solution \cite{Kitaev2006} and is independent of the sign of $K$. (b) and (c) show, respectively,  the MF specturm for the AFM and FM   Kitaev model  in the presence of magnetic field along the $[111]$ direction. Panels (d)-(f) show the MF spectrum for  $J\text{-}K\text{-}\Gamma$ model in the presence of magnetic field along the
      $[11\bar{2}]$ direction.
     Here we set $J=-0.05,$, $ \Gamma=0.2$ and $K=-1$. 
}
\label{Fig:spectrum}
\end{figure*}
\end{center}

\subsubsection{Mean-field spectrum}\label{Sec:MFspectrum}
Once the fields  $\mathbf{m}_L$ and $\mathrm{\Phi}^{a b}_{\delta}$ are self-consistently determined, ensuring satisfaction of constraints on average, we can proceed to analyze the mean-field solutions of the  extended  Kitaev model  in the presence of magnetic field.

In Fig.~\ref{Fig:spectrum}, we present the mean-field fermionic spectrum computed for the $J$-$K$-$\Gamma$ Hamiltonian for various values of $\mathbf{h}$ and $\kappa$. We begin by examining Fig.\ref{Fig:spectrum} (a), which illustrates the mean-field fermionic spectrum of the Kitaev model in the presence of the three-spin  term $\mathcal{H}_{\kappa}$, and  compare it  with the exact solution,
since this term  preserves the exact solvability of the Kitaev model.  
In particular, while the dispersionless gapped flux bands remain unchanged, a gap opens in the dispersive Majorana fermion spectrum at the Dirac $\mathbf{K}$-points, with its value consistent with perturbation theory.
The obtained phase is  the topologically gapped QSL phase with a Chern number of $\pm 1$ as predicted by Kitaev \cite{Kitaev2006}. 
 
 In Fig.~\ref{Fig:spectrum} (b) and (c), we explore the effect of the Zeeman term 
 $\mathcal{H}_{\mathcal{Z}}$ on the antiferromagnetic (AFM) and ferromagnetic (FM) Kitaev models, respectively. Here, the field 
$\mathbf{h}_{111}=h_{111} \hat{\mathbf{c}}$ 
 is applied  along the $\hat{\mathbf{c}}\propto[1,1,1]$ direction (refer to Fig. \ref{Fig:basic_honey_figure}) with $h_{111}=0.3$. In both cases,  the  ground state of the Kitaev model with the Zeeman field is  the chiral QSL with
  Chern number of $\pm 1$ \cite{Kitaev2006}.
The  Zeeman term opens the gap  in the Majorana spectrum at the Dirac $\mathbf{K}$-points,
and simultaneously lifts the three-fold degeneracy of the flat gapped  modes corresponding to the gauge fermions, causing them to acquire dispersion. 
  While this behavior is  observed   for both FM and AFM Kitaev interactions, the effect of the field is more pronounced in the case of the FM coupling.
The mean field spectrum of the Kitaev model with the magnetic field along the $\hat{\mathbf{a}}\propto[1,1,-2]$ direction  is found to exhibit qualitatively similar behavior to that seen in the $\mathbf{h}_{111}$ case.

\begin{center}
\begin{figure}
	\centering
	\includegraphics[width=1.\columnwidth]{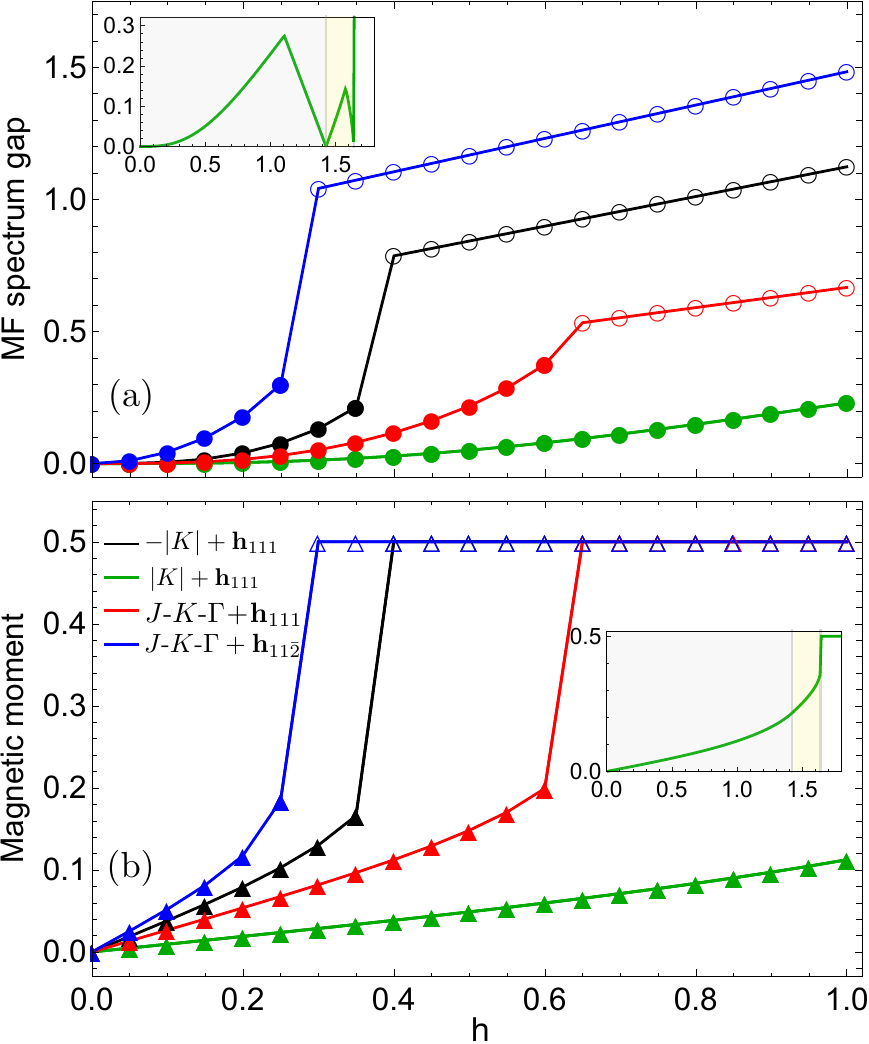}
     \caption{
     Evolution of the mean-field fermionic gap (a) and magnetic moment (b) with the strength of the Zeeman field $h$. Green and black curves represent the AFM and FM Kitaev models, respectively, with magnetic field in [111] direction. Red and blue curves correspond to the $J\text{-}K\text{-}\Gamma$ model with $K=-1$, $\Gamma=0.2$, and $J=-0.05$, with $\textbf{h}_{111}$ and $\mathbf{h}_{11\bar{2}}$, respectively. In (a), the plot shows gap at the Dirac $\mathbf{K}$-points for small fields (filled circles) shifting to $\mathbf{\Gamma}$-point for large fields (open circles). 
     The inset in (a) shows the mean-field gap for the AFM Kitaev model at larger field strengths, indicating two phase transitions with the first transition from the QSL to an intermediate QSL state at $h_{111}=1.4$, followed by second transition to polarized state at $h_{111}=1.65$.  In (b), filled triangles represent induced magnetization within the chiral QSL, while open triangles correspond to the fully polarized state. The inset in (b) shows magnetization for the AFM Kitaev model at larger field values, saturating after the transition from the intermediate QSL to the magnetically ordered state.}
\label{Fig:jkg_phase}
\end{figure}
\end{center}

   We  further illustrate the evolution of the Majorana fermion gap  and the 
   magnetization with respect to the applied field
    in Fig.~\ref{Fig:jkg_phase}.
 As shown in Fig.~\ref{Fig:jkg_phase} (a),  the gap, represented by the black curve for $K=-1$ and by
  green curve for  $K=1$,
  grows very slowly with increasing
 $h$, approximately following an $h^3$ behavior, consistent with the perturbative treatment by Kitaev~\cite{Kitaev2006}. It is worth noting that the antiferromagnetic (AFM) Kitaev model exhibits a slower, but still  $h^3$,  growth of the gap compared to the ferromagnetic (FM) case. Similar results were obtained by Ralko and Merino \cite{Ralko2020} (for  $K>0$ only), who attributed this correct behavior, even in the mean-field treatment, to the  explicit implementation of single-particle constraints using Lagrange multipliers.

As the magnetic field strength increases, the QSL state undergoes a transition to the polarized state, which is marked by a jump in the Majorana fermion's gap at 
$h_{111}=0.4$ 
for the FM model and at 
$h_{111}=1.65$
for the AFM model.
 Above these fields, the magnetic moments show saturation (Fig.~\ref{Fig:jkg_phase} (b)).
 This earlier transition to the polarized state in  the FM Kitaev model compared with the AFM case was previously    observed by the infinite density matrix
renormalization group (iDMRG) study \cite {Gohlkefield}.
The AFM Kitaev model  also shows an intermediate QSL at the field range between  
$h_{111}=1.4$ and $h_{111}=1.65$
\cite {Gohlkefield}. These two phase transitions can be seen in  the inset of  Fig.\ref{Fig:jkg_phase} (a). There, the mean-field gap  switches consequently from the $\mathbf{K}$-point to the $\mathbf{M}$-point and then to the $\mathbf{\Gamma}$-point of the Brillouin zone (BZ).

 In Fig.~\ref{Fig:spectrum} (d)-(f), we explore a mean-field spectrum for the $J$-$K$-$\Gamma$ model  relevant for
 $\alpha\text{-RuCl}_3$ \cite{sears2017phase,Ponomaryov2017, BalzPRB2019, kasahara_majorana_2018}. Specifically, we set $J=-0.05$, $ \Gamma=0.2$ and $K=-1$, which  allows  for a  QSL  mean-field ground state at zero field \cite{Susmita2023}. This parameter set shares a similar ratio of coupling strengths to that of 
  $\alpha\text{-RuCl}_3$.
   Note that further terms, specifically the third-neighbor Heisenberg coupling $J_3$ and a different off-diagonal $\Gamma'$ coupling stabilize the zigzag order in zero magnetic field in  $\alpha\text{-RuCl}_3$
\cite{JRau2014, Winter2017, MaksimovPRR2020, Song2016, Rousochatzakis2024}. However, in our current work we only consider the ideal lattice structure of the layered honeycomb materials without any distortions and therefore take $J\text{-}K\text{-}\Gamma$ as the minimal model. 
  The corresponding mean-field spectrum depicted in Fig.~\ref{Fig:spectrum} (d) retains a Dirac cone at the $\mathbf{K}$-points, while the flux bands are already dispersing due to the loss of commutation of the flux operators with the $\mathcal{H}_0$ Hamiltonian. 
 
 In Fig.~\ref{Fig:spectrum} (e) and (f), the mean-field spectrum obtained from reorienting the field to the in-plane direction $\mathbf{h}_{11\bar{2}}=h_{11\bar{2}} \hat{\mathbf{a}}$, along the $\hat{\mathbf{a}}=[1,1,-2]$,  is shown for
$h_{11\bar{2}}=0.1$ and $0.2$, respectively. The Dirac points again gap out under this field as well, and this gap  is larger  for larger strength of the magnetic field.

As in the case of the pure Kitaev model in the field,  the  QSL  of the $J$-$K$-$\Gamma$ model  also
undergoes a transition to the polarized state. The critical value at which this transition occurs depends on the field direction, with the critical  magnetic field along ${\hat{\mathbf{a}}}$ direction  $h^*_{11\bar{2}}=0.3$  being lower than the one in the field along the $\hat{\mathbf{c}}$ direction, $h^*_{111}=0.65$. 
Similar field direction dependence of the critical field has been observed in
high field measurements of the magnetization in $\alpha\text{-}\text{RuCl}_3$~\cite{Zhou2023}.

\subsection{Free phonons in 2D elastic medium}
We provide a brief overview of the dynamics of acoustic phonons on the honeycomb lattice, as discussed in Ref.~\cite{Mengxing2020}. Acoustic phonons can be described by a long-wavelength effective action, involving displacement fields $\bm{u}$ representing atom displacements. For systems with TRS, the lowest-order action is:
\begin{equation}\label{TRSaction}
S_{ph}^{(s)}=\int d^2xd\tau[\rho (d_\tau\bm{u})^2+\frac{1}{2}C_{ijlk}\epsilon_{ij}\epsilon_{lk}],
\end{equation}
where $\rho$ is the mass density of lattice ions, and $C_{ijlk}$ is the  elastic modulus tensor with only two independent nonzero coefficients  due to lattice symmetries 
 and $\epsilon_{ij}$ is the strain tensor.
The elastic energy term for in-plane phonon modes in Eq.~(\ref{TRSaction}) can be written in terms of $\epsilon_{aa}+\epsilon_{bb}$ and ${\epsilon_{aa}-\epsilon_{bb},\,2 \epsilon_{ab}}$ with 
  $\hat{\mathbf{a}}$ and $\hat{\mathbf{b}}$  being defined in  Fig.\ref{Fig:basic_honey_figure}, forming the basis for irreducible representations $A_{1g}^{\text{ph}}$ and $E_g^{\text{ph}}$ of the $D_{3d}$ point group, respectively \cite{Mengxing2020}.

After performing a Fourier transform to momentum and  frequency space and diagonalizing, we obtain the longitudinal ($\mu=\parallel$) and transverse ($\mu=\perp$) components of the acoustic phonon spectrum $\Omega_{\textbf{q}}^{\mu}$ and the polarization vectors $\hat{\textbf{e}}_{\textbf{q}}^{\mu}$, given by
\begin{eqnarray}
\begin{array}{cl}
\Omega_{\textbf{q}}^{\mu}=v_{s}^{\mu}q, & \hat{\textbf{e}}_{\textbf{q}}^{\parallel}=\{\cos\theta_{\textbf{q}},\sin\theta_{\textbf{q}}\}, \\
\textbf{u}_\textbf{q}=\sum_{\mu}\hat{\textbf{e}}^{\mu}_{\textbf{q}}\tilde{u}^{\mu}_{\textbf{q}}, & \hat{\textbf{e}}_{\textbf{q}}^{\perp}=\{-\sin\theta_{\textbf{q}},\cos\theta_{\textbf{q}}\},
    \end{array}
\end{eqnarray}
where where $\tilde{u}^{\mu}_{\textbf{q}}$denotes the magnitude of lattice displacement
in the $\mu$  polarization,
$q=\sqrt{q_a^2+q_b^2}$, and $\theta_{\textbf{q}}$ is the angle between the phonon momentum $\textbf{q}$ and the 
 orthorhombic $\hat{\textbf{a}}$ direction (see Fig.~\ref{Fig:basic_honey_figure}).

While the phonons do not directly couple to TRS-breaking terms like magnetic fields, the magneto-elastic coupling, discussed in the next section, couples spin and lattice degrees of freedom. This indirect interaction channels magnetic field effects, originating from the Zeeman and $\kappa$ terms in the spin Hamiltonian, into the phonons. Consequently, the phonon effective action acquires a contribution from the Hall viscosity \cite{Avron1995, Barkeshli2012}:
\begin{equation}\label{eq:antisymmetric_action}
S_{ph}^{(a)}=\int d^2xdt\eta_{ijlk}^{(a)}\epsilon_{ij}\dot{\epsilon}_{lk},
\end{equation}
where the viscosity tensor is antisymmetric, i.e., $\eta_{ijlk}^{(a)}=-\eta_{lkij}^{(a)}$. Symmetry constraints reduce the number of independent components in the Hall viscosity to just one, denoted as  $\eta_H$ \cite{Mengxing2020}.

\subsection{Magneto-elastic coupling}\label{sec:magneto-elast}

To study the impact of Majorana excitations of the Kitaev QSL on phonon dynamics, we also need to compute the Majorana fermion-phonon coupling vertices \cite{Mengxing2020}. This coupling arises from the dependence of exchange couplings $\mathcal{J}=J, K,\Gamma$ on the relative positions between spins. Assuming the coupling strengths depend only on the distance $r$ between atoms and using the adiabatic approximation where the electronic state forming the local spin moments responds instantaneously to small movements of the sites, we express the coupling term to leading order as $\mathcal{J}(r)=\mathcal{J}_{\text{eq}}+(\frac{d \mathcal{J}(r)}{d r})_{\text{eq}}{\delta r}$, where the spin-phonon coupling strength is $\lambda_{\mathcal{J}}=(\frac{d\mathcal{J}(r)}{d r})_{\text{eq}}$.

To determine the full set of  independent magneto-elastic couplings, we consider symmetry   irreducible representations (irreps). The Hamiltonian's symmetry depends on the magnetic field's direction in the Zeeman term $H_{\mathcal{Z}}$. Thus, we will analyze two magnetic field orientations, $h$, in the subsequent subsections.
 
\begin{table}[t]
\begin{center}
\begin{tabular}{| c| c|  }
\hline
\, & $A_{1g}$ \\ 
\hline
$K$ &$\Xi_{\bm{r},K}^{A_1g}=\sigma_{\textbf{r}}^x\sigma_{\textbf{r}+\bm{\delta}_x}^x+\sigma_{\textbf{r}}^y\sigma_{\textbf{r}+\bm{\delta}_y}^y+\sigma_{\textbf{r}}^z\sigma_{\textbf{r}+\bm{\delta}_z}^z$ \\  
\hline
$J$&$\Xi_{\bm{r},J}^{A_1g}=\sigma_{\textbf{r}}^x\sigma_{\textbf{r}+\bm{\delta}_{y,z}}^x+\sigma_{\textbf{r}}^y\sigma_{\textbf{r}+\bm{\delta}_{x,z}}^y+\sigma_{\textbf{r}}^z\sigma_{\textbf{r}+\bm{\delta}_{x,y}}^z$ \\
\hline  
$\Gamma$&$\Xi_{\bm{r},\Gamma}^{A_1g}=\sigma_{\textbf{r}}^y\sigma_{\textbf{r}+\bm{\delta}_{x}}^z+\sigma_{\textbf{r}}^z\sigma_{\textbf{r}+\bm{\delta}_{x}}^y+\sigma_{\textbf{r}}^x\sigma_{\textbf{r}+\bm{\delta}_{y}}^z$ \\
&$+\sigma_{\textbf{r}}^z\sigma_{\textbf{r}+\bm{\delta}_{y}}^x+\sigma_{\textbf{r}}^x\sigma_{\textbf{r}+\bm{\delta}_{z}}^y+\sigma_{\textbf{r}}^y\sigma_{\textbf{r}+\bm{\delta}_{z}}^x$ \\  
\hline
strain &$\epsilon_{aa}+\epsilon_{bb}$\\
\hline
\, & $E_g$\\ 
\hline
$K$  & $(\Xi_{\bm{r},K}^{E_{g}^1},\Xi_{\bm{r},K}^{E_{g}^2})=(\sigma_{\textbf{r}}^x\sigma_{\textbf{r}+\bm{\delta}_x}^x+\sigma_{\textbf{r}}^y\sigma_{\textbf{r}+\bm{\delta}_y}^y-2\sigma_{\textbf{r}}^z\sigma_{\textbf{r}+\bm{\delta}_z}^z,$\\
&$\sqrt{3}(\sigma_{\textbf{r}}^x\sigma_{\textbf{r}+\bm{\delta}_x}^x-\sigma_{\textbf{r}}^y\sigma_{\textbf{r}+\bm{\delta}_y}^y))$ \\  
\hline
$J$ & $(\Xi_{\bm{r},J}^{E_{g}^1},\Xi_{\bm{r},J}^{E_{g}^2})=(\sigma_{\textbf{r}}^x\sigma_{\textbf{r}+\bm{\delta}_{y,z}}^x+\sigma_{\textbf{r}}^y\sigma_{\textbf{r}+\bm{\delta}_{x,z}}^y-2\sigma_{\textbf{r}}^z\sigma_{\textbf{r}+\bm{\delta}_{x,y}}^z,$\\
&$\sqrt{3}(\sigma_{\textbf{r}}^x\sigma_{\textbf{r}+\bm{\delta}_{y,z}}^x-\sigma_{\textbf{r}}^y\sigma_{\textbf{r}+\bm{\delta}_{x,z}}^y))$ \\  
\hline  
$\Gamma$ & $(\Xi_{\bm{r},\Gamma}^{E_{g}^1},\Xi_{\bm{r},\Gamma}^{E_{g}^2})=(\sigma_{\textbf{r}}^y\sigma_{\textbf{r}+\bm{\delta}_{x}}^z+\sigma_{\textbf{r}}^z\sigma_{\textbf{r}+\bm{\delta}_{x}}^y+\sigma_{\textbf{r}}^x\sigma_{\textbf{r}+\bm{\delta}_{y}}^z$ \\
&$+\sigma_{\textbf{r}}^z\sigma_{\textbf{r}+\bm{\delta}_{y}}^x-2(\sigma_{\textbf{r}}^x\sigma_{\textbf{r}+\bm{\delta}_{z}}^y+\sigma_{\textbf{r}}^y\sigma_{\textbf{r}+\bm{\delta}_{z}}^x),$ \\  &$\sqrt{3}\left[\sigma_{\textbf{r}}^y\sigma_{\textbf{r}+\bm{\delta}_{x}}^z+\sigma_{\textbf{r}}^z\sigma_{\textbf{r}+\bm{\delta}_{x}}^y-(\sigma_{\textbf{r}}^x\sigma_{\textbf{r}+\bm{\delta}_{y}}^z+\sigma_{\textbf{r}}^z\sigma_{\textbf{r}+\bm{\delta}_{y}}^x)\right])$ \\  
\hline
strain &$\bigl(\epsilon_{aa}-\epsilon_{bb},2\epsilon_{ab}\bigr)$\\
\hline  
\end{tabular}
\caption{Basis functions of spins and phonons in the $A_{1g}$ and $E_g$ irreps of the $D_{3d}$ point group. 
}
\label{table:irrA1g}
\end{center}
\end{table}
\subsubsection{Magnetic field along  the $[1,1,1]$ direction}
For the  zero-field and the field along [1,1,1] direction,  the point group is isomorphic  to $D_{3d}$ and 
the magneto-elastic
Hamiltonian has two independent symmetry channels, $A_{1g}$ and  $E_g$,
\begin{equation}\label{eq:hc_d3d}
    \mathcal{H}_c=\mathcal{H}_c^{A_{1g}}+\mathcal{H}_c^{E_g},
\end{equation}
where
\begin{equation}\label{eq:hc_d3d_a1g}
    \mathcal{H}_c^{A_{1g}}=\lambda_{A_{1g}}\sum_{\bm{r},\mathcal{J}}(\epsilon_{aa}+\epsilon_{bb})\Xi^{A_{1g}}_{\bm{r},\mathcal{J}},
\end{equation}
\begin{equation}\label{eq:hc_d3d_eg}  \mathcal{H}_c^{E_{g}}=\lambda_{E_{g}}\sum_{\bm{r},\mathcal{J}}\{(\epsilon_{aa}-\epsilon_{bb})\Xi^{E_{g}^1}_{\bm{r},\mathcal{J}}+2\epsilon_{ab}\Xi^{E_{g}^2}_{\bm{r},\mathcal{J}}\},
\end{equation}
and $E_{g}^1$ and $E_{g}^2$ are the two basis functions of
the two dimensional  $E_g$ representation, and   $\lambda_{A_{1g}}$ , $\lambda_{E_g}$ are 
the coupling constants.
 Also, for convenience, we have changed the notations  for sites as $i\in A \rightarrow {\bf r},\, j\in B \rightarrow {\bf r}+\boldsymbol\delta$.
In the following, we set $\lambda_{A_{1g}}$=$\lambda_{E_g}$=$\lambda$. The explicit expressions for the spin space basis functions $\Xi^{A_{1g}}_{\bm{r},\mathcal{J}}$, $\Xi^{E_{g}^1}_{\bm{r},\mathcal{J}}$ and $\Xi^{E_{g}^2}_{\bm{r},\mathcal{J}}$ of the $D_{3d}$ point group can be found in Table~\ref{table:irrA1g}.

\subsubsection{Magnetic field along  the $[1,1,\bar{2}]$ direction}

When the field is applied along $\hat{\bf a}=[1,1,\bar{2}]$, the $D_{3d}$ symmetry is reduced to $C_{2h}$. Accordingly the magnetoelastic coupling is then written as
\begin{equation}\label{eq:hc_c2h}
    \mathcal{H}_c=\mathcal{H}_c^{A_{g}}+\mathcal{H}_c^{B_g},
\end{equation}
with
\begin{equation}\label{eq:hc_c2h_ag}
    \mathcal{H}_c^{A_{g}}=\lambda\sum_{\bm{r},\mathcal{J}}\{(\epsilon_{aa}+\epsilon_{bb})\Xi^{A_{1g}}_{\bm{r},\mathcal{J}}+(\epsilon_{aa}-\epsilon_{bb})\Xi^{E_{g}^1}_{\bm{r},\mathcal{J}}\},
\end{equation}

\begin{equation}\label{eq:hc_c2h_bg}
    \mathcal{H}_c^{B_{g}}=\lambda\sum_{\bm{r},\mathcal{J}}2\epsilon_{ab}\Xi^{E_{g}^2}_{\bm{r},\mathcal{J}}.
\end{equation}
Here the use of the same basis functions $\Xi_{\bm{r},\mathcal{J}}$ demonstrates the change in irreducible representations (irreps)  channels when lowering the symmetry from $D_{3d}$ to $C_{2h}$. Specifically, the $A_{1g}$ becomes $A_g$ and the two-dimensional $E_g$  representation breaks into two one-dimensional irreps --  $E_{g}^1$ transforming as $A_g$, and $E_{g}^2$ transforming as $B_g$. This change can also be verified using the character tables of the two point groups.

\subsubsection{Mean-field coupling vertex}
Using the Majorana mean-field solution in  the spin-phonon coupling Hamiltonian (Eqs. \eqref{eq:hc_d3d}-\eqref{eq:hc_d3d_eg} and Eq. \eqref{eq:hc_c2h}-\eqref{eq:hc_c2h_bg}) and expressing the phonon modes in terms of the longitudinal and transverse eigenmodes, we can write
\begin{equation}
\begin{split}
\mathcal{H}_{\textbf{q},\textbf{k}}^{\parallel}=&\tilde{\textbf{u}}_{\textbf{q}}^{\parallel}{\mathcal C}_{-\textbf{k}-\textbf{q}}^T\hat{\lambda}_{\textbf{q},\textbf{k}}^{\parallel}{\mathcal C}_{\textbf{k}},\\
\mathcal{H}_{\textbf{q},\textbf{k}}^{\perp}=&\tilde{\textbf{u}}_{\textbf{q}}^{\perp}{\mathcal C}_{-\textbf{k}-\textbf{q}}^T\hat{\lambda}_{\textbf{q},\textbf{k}}^{\perp}{\mathcal C}_{\textbf{k}},
\end{split}
\end{equation}
where $\hat{\lambda}_{\textbf{q},\textbf{k}}^{\mu}$ are the spin-phonon coupling vertices in the Majorana basis  (see further details in App.~\ref{appx:magnetoelastic_details}). For calculating the phonon polarization bubble, we write the coupling Hamiltonian in the basis of the Bogoliubov quasiparticles as,
\begin{equation}
\begin{array}{cc}
  \mathcal{H}_{\textbf{q},\textbf{k}}^{\mu}=\tilde{u}_{\textbf{q}}^{\mu}{\mathcal B}^{\dagger}_{\textbf{k}+\textbf{q}}\tilde{\lambda}_{\textbf{q},\textbf{k}}^{\mu}{\mathcal B}_{\textbf{k}},  \\[0.2cm]
    \tilde{\lambda}_{\textbf{q},\textbf{k}}^{\mu}=V_{\textbf{k}+\textbf{q}}^{\dagger}\lambda_{\textbf{q},\textbf{k}}^{\mu}V_{\textbf{k}},
\end{array}
\end{equation}
where $\mu=\parallel,\ \perp$ and  the columns in  $V_{\bm{k}}$ 
contain  the eigenstates in terms of the Majorana basis
(See Eq.~\eqref{eq:diagonalizng_transformations}, \eqref{eq:diagonalizng_transformations1} in App.~\ref{appx:peter_you_better_get_this_appendix_done_already}). The coupling vertices are divided into four blocks according to the division into creation and annihilation sectors (see App.~\ref{appx:magnetoelastic_details}):
\begin{equation}
\tilde{\lambda}_{\textbf{q},\textbf{k}}^{\mu}=\left[\begin{array}{cc}
\tilde{\lambda}_{\textbf{q},\textbf{k},11}^{\mu}&\tilde{\lambda}_{\textbf{q},\textbf{k},12}^{\mu}\\
\tilde{\lambda}_{\textbf{q},\textbf{k},21}^{\mu}&\tilde{\lambda}_{\textbf{q},\textbf{k},22}^{\mu}
\end{array}\right].
\end{equation}

The corrections to the acoustic phonon effective action due to the magnetoelastic coupling is calculated through the phonon polarization bubble to one loop order, given by
\begin{equation}
\begin{split}
    \Pi^{\mu\nu}(\textbf{q},\tau)=&\langle T_{\tau}(\mathcal{B}^{\dagger}_{\textbf{k}+\textbf{q}}\tilde{\lambda}_{\textbf{q},\textbf{k}}^{\mu}\mathcal{B}_{\textbf{k}})(\tau)(\mathcal{B}^{\dagger}_{\textbf{k}'-\textbf{q}}\tilde{\lambda}_{-\textbf{q},\textbf{k}'}^{\mu}\mathcal{B}_{\textbf{k}'})(0)\rangle.
\end {split}
\end{equation}
The details of the calculation of the phonon polarization bubble are given in App.~\ref{appx:bubble_calculation}. 

\section{Results}

In this section, we present our results for the dynamics of acoustic phonons coupled to a chiral spin liquid in a Kitaev magnet under an external magnetic field.

\begin{center}
\begin{figure*}
	\centering
	\includegraphics[width=1.0\linewidth]{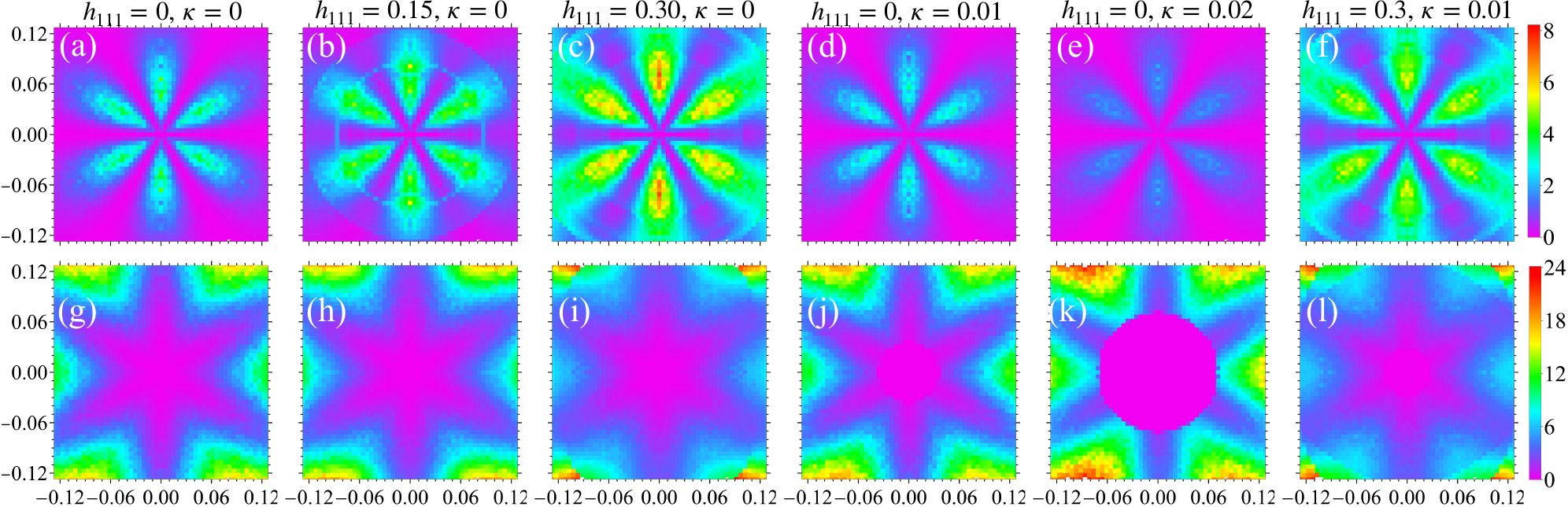}
     \caption{
The sound-attenuation coefficient $\alpha_s^{\parallel}(\textbf{q})$ for the longitudinal phonon mode computed  at $T=0.1$ for the  AFM Kitaev model ($K=1$) in the presence of the Zeeman field $h_{111}$ and $\kappa$ term. Panels (a)-(f) display results for the ph-dominant regime, $v_s < v_F$, while panels (g)-(l) show the pp-dominant regime, $v_s > v_F$.  
Only the ph-contribution for $v_s < v_F$ and the pp-contribution for $v_s > v_F$ from the $E_g$ irrep  are plotted. Phonon momentum $\textbf{q}$ is in the region $(q_a,q_b)\in [-0.12\pi,0.12\pi]^2$.  The sound velocity is set to $v_s=0.3$ for panels (a)-(f) and $v_s=3.5$ for panels (g)-(l). The Fermi velocity,  $v_F=3$, is  determined from the slope of the Dirac cone of the Kitaev spectrum. The artificial energy broadening is $\delta=0.2$. All energies are in the units of $|K|$.
}
\label{Fig:alon_khkappa_grid}
\end{figure*}
\end{center}

\subsection{Sound attenuation}
We first compute the sound attenuation, which measures the decay of the phonon amplitude with distance away from the driving source due to scattering processes. This attenuation is described by the lossy acoustic phonon wavefunction:
\begin{equation}
    \mathbf{u(\textbf{r},t)}=\mathbf{u}_0\mathrm{e}^{-\alpha_s(\textbf{q})r}\mathrm{e}^{i(\Omega t-\mathbf{q}\cdot\mathbf{r})},
\end{equation}
where $ \mathbf{u(\textbf{r},t)}$ is the lattice displacement vector, $\mathbf{u}_0=\mathbf{u}(\textbf{r},t=0)$, $\Omega$ is the acoustic wave frequency, and $\textbf{q}$ is the propagation vector. Since the sound attenuation coefficient can differ for different acoustic modes, we denote
 $\alpha^{\mu}_s(\textbf{q})$  as  the sound attenuation coefficient for the two phonon polarizations $\mu=\parallel,\perp$. It
can be calculated from the imaginary part of the phonon polarisation bubble as \cite{Mengxing2020}
\begin{equation}
    \alpha_s^{\mu}(\textbf{q})=-\frac{1}{v_s^2q}\mathrm{Im}[\Pi^{\mu\mu}(\textbf{q},\Omega)]_{\Omega=v_sq}.
\end{equation}

The attenuation of acoustic phonons coupled to the Kitaev QSL arises 
from 
phonon decay into fermions and from phonon scattering off fermions, as is allowed by the magneto-elastic coupling.
In the pure Kitaev model at low temperatures (below the flux gap), only the low-energy Majorana fermion states near the Dirac cones significantly contribute to these scattering processes \cite{Mengxing2020}. The phase space for these processes is determined by the relative  magnitudes of the Fermi velocity, $v_F$, and the phonon sound velocity, $v_s$, constrained by momentum and energy conservation. In the pure Kitaev model, when $v_s < v_F$, only particle-hole (ph) processes, where a phonon scatters a fermion from an occupied state to a higher energy unoccupied state, are allowed \cite{Mengxing2020}. Conversely, when $v_s > v_F$, only particle-particle (pp) processes, where a phonon decays into a pair of fermions, are allowed.

When interactions beyond the Kitaev are  present in the spin Hamiltonian, 
the mean-field fermionic spectrum may allow for pp-processes even in the regime $v_s<v_F$.
In this case, the modified mean-field spectrum, in  which the originally flat flux bands start to disperse, may allow the kinematic constraints to be satisfied for exciting two fermions away from the Dirac cones (e.g., near the $\mathbf{\Gamma}$-point of the BZ). However, despite this possibility, the particle-hole (ph) processes remain dominant \cite{Susmita2023}.

 The presence of the Zeeman and $\kappa$ terms further modifies the Majorana fermion spectrum, leading to observable changes in sound attenuation. We will next explore these effects.

 \begin{center}
\begin{figure*}
	\centering
	\includegraphics[width=1.0\linewidth]{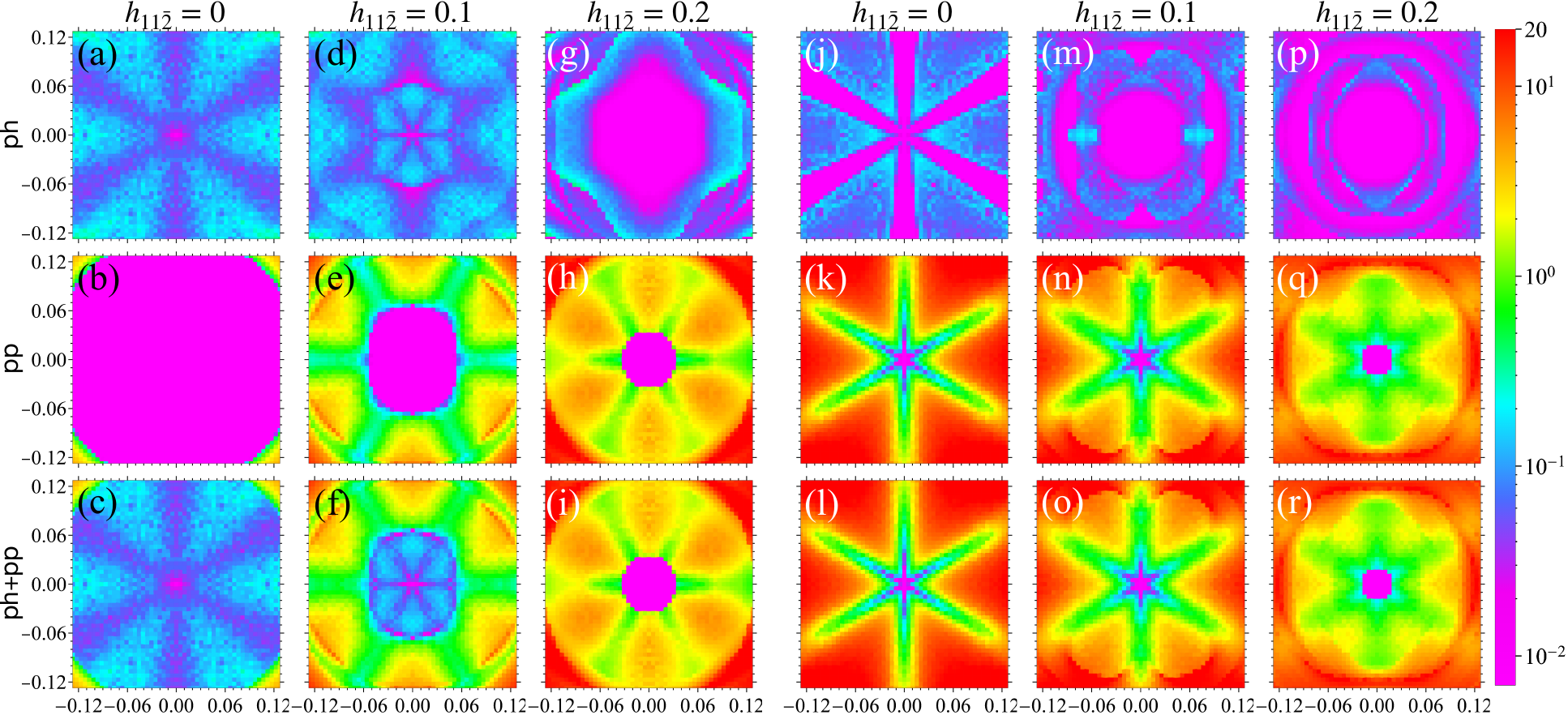}
     \caption{
   The transverse sound attenuation coefficient, $\alpha_s^{\perp}(\textbf{q})$,  and the  longitudinal one, $\alpha_s^{\parallel}(\textbf{q})$,  computed for the $J\text{-}K\text{-}\Gamma$ model in an  in-plane field $\textbf{h}_{11\bar{2}}$ at $T=0.05$ are shown in (a)-(i) and  (j)-(r), respectively. Here, $K=-1$, $J=-0.05$, and $\Gamma=0.2$.  We display contributions from the ph-process [(a), (d), (g); (j), (m), (p)], pp-process [(b), (e), (h); (k), (n), (q)], and their sum [(c), (f), (i); (l), (o), (r)], originating from the  $E_g$ symmetry channel. The phonon momentum $\textbf{q}$ belongs to the region $(q_a,q_b)\in [-0.12\pi,0.12\pi]^2$. For (a)-(i), $v_s=2.7$; for (j)-(r), $v_s=3.3$; and $v_F=3$. The artificial  energy broadening is $\delta=0.2$. All energies are in the units of $|K|$. 
}
\label{Fig:attn_jkg_grid}
\end{figure*}
\end{center}

\subsubsection{Sound attenuation for pure Kitaev model with field}

 We start  with the pure Kitaev model in the presence of TRS breaking  terms, $\textbf{h}_{111}$ and $\kappa$, denoted as the $K+\textbf{h}_{111}+\kappa$ model. 
 
Fig.~\ref{Fig:alon_khkappa_grid} displays the $\textbf{q}$ dependence of the longitudinal sound attenuation coefficient, $\alpha_s^{\parallel}(\textbf{q})$, for different combinations of $\textbf{h}_{111}$ and $\kappa$.
Figs.~\ref{Fig:alon_khkappa_grid} (a) and (g) display the results for the pure Kitaev model, serving as a reference point for studying the effects of these terms.
Panels Fig.~\ref{Fig:alon_khkappa_grid} (a)-(f) (top row) correspond to the ph-dominant regime with $v_s=0.3$, while panels Fig.~\ref{Fig:alon_khkappa_grid} (g)-(l) (bottom row) correspond to the pp-dominant regime with $v_s=3.5$.

 In the absence of the $\kappa$-term, the Zeeman field $\textbf{h}_{111}$ primarily increases the overall magnitude of the sound attenuation for the 
 ph-processes (Fig.~\ref{Fig:alon_khkappa_grid} (b), (c)) and only slightly decreases it for the pp-processes (Fig.~\ref{Fig:alon_khkappa_grid}(h), (i)), compared to the pure case.
This behavior is due to $\textbf{h}_{111}$ opening a very small gap in the Majorana fermion spectrum while adding significant dispersion to the flat bands 
(see Fig.~\ref{Fig:spectrum}),  leading to an increase in the low-energy density of states (DOS).
Consequently, we observe an increase in the intensity of the ph-processes and a decrease in the pp-processes. 
The Zeeman term also introduces extra features in the six-fold shape of the sound attenuation, arising from the dispersion of the flat bands in the Majorana fermion spectrum (see Fig.~\ref{Fig:spectrum}).

 In contrast, a non-zero $\kappa$ term decreases the magnitude for the ph-process Fig.\ref{Fig:alon_khkappa_grid} (d), (e) and increases it for the pp-process Fig.\ref{Fig:alon_khkappa_grid} (j), (k).  Most notably,  a non-zero $\kappa$ term results in a disk-shaped region centered around $q=0$ with no sound attenuation from the pp-process seen in Fig.\ref{Fig:alon_khkappa_grid} (j), (k), and this disk's radius increases with increasing $\kappa$.  This behavior is again related to the modified Majoarana fermion pectrum. 
  The $\kappa$ term opens a substantial gap in the fermionic spectrum at the $\mathbf{K}$-point of the BZ, while leaving the flat flux bands unchanged (see Fig.~\ref{Fig:spectrum} (a)). This gap opening decreases the low-energy DOS and increases the higher energy DOS. 
 The decrease in the low-energy DOS reduces the phase space for the ph-process, which requires occupied low-energy states. However, the increased DOS for higher energies results in more unfilled fermionic states at low temperatures, leading to an increase in the pp-process. This increase is observed only above a finite value of the phonon momentum $q$, above which the energy constraint can be satisfied. As the gap size increases with $\kappa$, higher energy phonons are needed to 
 populate two fermionic states at or above the gap for larger $\kappa$. Therefore, the radius of the region without the pp-process increases with $\kappa$.  A combination of the effects of $\textbf{h}_{111}$ and $\kappa$ is shown in Fig.~\ref{Fig:alon_khkappa_grid} (f), (l).

\subsubsection{Sound attenuation in the $J\text{-}K\text{-}\Gamma$ model with the in-plane magnetic field}

 Recent ultrasound studies of the Kitaev candidate material $\alpha\text{-}\text{RuCl}_3$  \cite{Hauspurg2024}
 have provided  insights into the temperature and field dependence of sound attenuation for various acoustic modes.
 Specifically,  both ultrasound~\cite{Hauspurg2024} and inelastic x-ray scattering~\cite{Haoxiang2021} measurements  in $\alpha\text{-RuCl}_3$ revealed that the velocity of the in-plane transverse acoustic phonons is approximately $v_{s}^{\perp}\sim 16 \text{ meV \AA}$,  while the velocity of the in-plane longitudinal acoustic phonons is approximately $v_{s}^{\parallel}\sim 20 \text{ meV \AA}$.
 The Fermi velocity of the Majorana fermions in $\alpha\text{-RuCl}_3$  is estimated to be about
  $v_{F} \sim 18\, \text{ meV\AA}$ \cite{Miao2020,LebertPRB2022,Haoxiang2021}. 
  (Assuming  $K\sim-6.0$ $\text{meV}$ ~\cite{MaksimovPRR2020}, $18\, \text{ meV\AA}$ 
   corresponds to  $v_F=3$ in units of $|K|{\text \AA}$.)
   This puts $\alpha-$RuCl$_3$ in the regime $v_s^{\perp} < v_F < v_s^{\parallel}$, allowing for the exploration of both sound attenuation channels within the same compound. 

The ultrasound experiment suggested that in $\alpha\text{-}\text{RuCl}_3$, the sound attenuation behavior over a wide temperature range above the Néel ordering temperature $T_N = 7\text{K}$ can be interpreted as a signature of the fractionalized excitations of a Kitaev magnet with comparable Fermi and sound velocities. Transverse sound modes, which have lower velocities, scatter off particle-hole (ph) excitations, inducing a characteristic $T$-linear damping, while longitudinal modes with higher velocities tap into the fermionic particle-particle (pp) continuum. This finding provides further support for the hypothesis that $\alpha\text{-}\text{RuCl}_3$ is in proximity to the Kitaev spin liquid regime.

Before presenting our results close to the experimental conditions, we briefly discuss the transverse sound attenuation coefficient $\alpha^{\perp}_s(\textbf{q})$ for the $J\text{-}K\text{-}\Gamma$ model (model parameters $J=-0.05, K=-1, \Gamma=0.2$) with $v_s=2$ and no external magnetic field.  In this case (shown below), we observe the clear six-fold petal shape of the sound attenuation  in the ph-channel (a), while the pp-channel is completely absent (b):
\begin{equation*}
     \includegraphics[width=1\columnwidth]{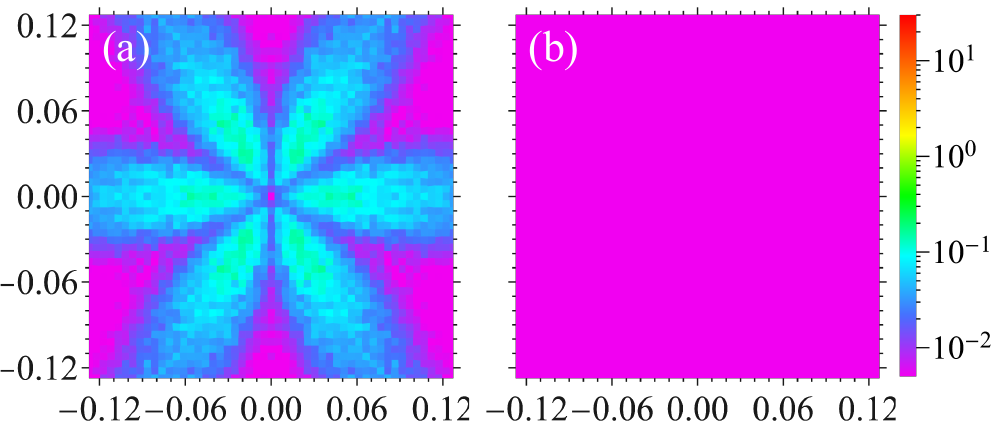}
 \end{equation*}
   This is because, in the absence of a magnetic field, the Dirac cones remain intact and contribute to the ph-processes for phonon velocities smaller than the Fermi velocity, even in the presence of non-Kitaev interactions $J$ and $\Gamma$.

 We now analyze the behavior of sound attenuation in the $J$-$K$-$\Gamma$ model with an in-plane magnetic field $\textbf{h}_{11\bar{2}}$. 
 We again set the exchange couplings  to $J=-0.05$, $K=-1$, $\Gamma=0.2$, but put  $v_s=2.7  <v_F$ for the transverse mode and $v_s=3.3>v_F$ for the longitudinal mode (for simplicity, we  now omit the units for the velocities, assuming they are in $|K|{\text \AA}$).
This set of parameters closely approximates the parameters describing $\alpha\text{-}\text{RuCl}_3$.  
The resulting sound attenuation is shown in Fig.~\ref{Fig:attn_jkg_grid}, where the attenuation for longitudinal and transverse polarized phonons is rotated by 90$^\circ$ \cite{Mengxing2020}.

As expected, for $v_s=2.7$ and $h_{11\bar{2}}=0$, only the ph-processes contribute to the sound attenuation, displaying the  "flower" shape in 
Fig.~\ref{Fig:attn_jkg_grid} (c), originated entirely from   the ph-processes  (see Fig.~\ref{Fig:attn_jkg_grid} (a)) since  the pp-processes (see Fig.~\ref{Fig:attn_jkg_grid}(b)) are nearly absent. 
 With increasing  magnetic field, at $h_{11\bar{2}}=0.1$ and $h_{11\bar{2}}=0.2$,
shown in Fig.~\ref{Fig:attn_jkg_grid} (d)-(i), the ph-processes dominate for small $q$, whereas the pp-processes become dominant for larger $q$. Conversely, for $v_s=3.3$, the pp-processes dominate for both zero and finite Zeeman fields, as shown in Fig.~\ref{Fig:attn_jkg_grid} (j)-(r).  These changes are due to the modified Majorana fermion spectrum, which, in the presence of the field, includes dispersive vison modes and wider Dirac cones with  modified  Fermi velocity $\tilde{v}_F < v_F$. The latter shifts the balance between the sound and Fermi velocities, allowing kinematic constraints to be satisfied for the pp-processes. 

Additionally, a non-zero Zeeman field $\textbf{h}_{11\bar{2}}$ reduces the six-fold symmetry of the sound attenuation pattern to two-fold symmetry.
Namely, when $h_{11\bar{2}}\neq0$,  only horizontal mirror plane symmetry remains in the petals, consistent with the surviving $\mathcal{T}C_{2Z}$ symmetry under the $[1,1,-2]$ field (see symmetry Sec.~\ref{sec:symmetry}).

\begin{figure}
	\centering
	\includegraphics[width=0.98\columnwidth]{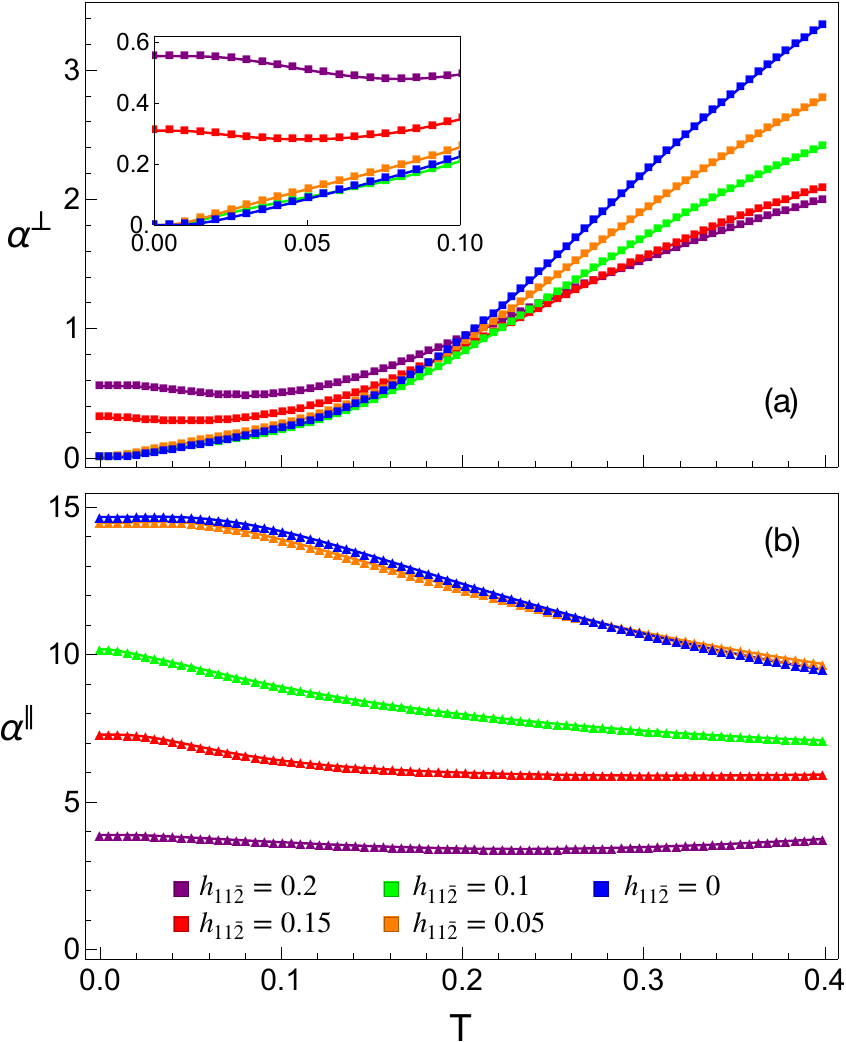}
     \caption{
Temperature evolution of the sound attenuation coefficient for the $J$-$K$-$\Gamma$ model with in-plane field $\mathbf{h}_{11\bar{2}}$: (a) transverse component, $\alpha_s^{\perp}(\textbf{q})$, for $(v_s=2.7)<(v_F=3)$; (b) longitudinal component, $\alpha_s^{\parallel}(\textbf{q})$, for $(v_s=3.3)>(v_F=3)$.  Parameters of the model are $K=-1$, $J=-0.05$, $\Gamma=0.2$. The plots show the sum of ph- and pp-processes from the original $E_g$ irreducible representation. The phonon momentum for (a) and (b) was taken to be $\bm{q}=(0.05\pi,0)$, and $\bm{q}=(0.1\pi,0)$, respectively. 
Artificial energy broadening is set $\delta=0.2$. All energies are in the units of $|K|$.
     }
\label{Fig:alpha_jkg_temp_evolution}
\end{figure}

Next, let us discuss the  temperature dependence of sound attenuation
in the $J$-$K$-$\Gamma$  model presented in Fig.~\ref{Fig:alpha_jkg_temp_evolution}. In panel (a), we observe a linear in $T$ dependence of
$\alpha_s^{\perp}(\textbf{q})$ computed  for $v_s=2.7$ and $h_{11\bar{2}}=0.0$, 0.05 and 0.1.  However for 
$h_{11\bar{2}}=0.15, 0.2$, the sound attenuation becomes temperature-independent due to the pp-processes. This behavior arises from the field-modified Majorana fermion spectrum and consequent changes in the kinematic conditions. 
In Fig.~\ref{Fig:alpha_jkg_temp_evolution}  (b) we show the temperature evolution of the longitudinal component $\alpha_s^{\parallel}$ with $v_s=3.3$. Here, pp-processes dominate at all values of $h_{11\bar{2}}$. 
However, because the gap increases with the field, the sound attenuation becomes smaller at low temperatures, since the energy of the phonon is insufficient  for creating two fermions, i.e., the kinematic constraints cannot be satisfied. With increasing temperature, we see a slow decrease of $\alpha_s^{\parallel}$. 

To summarize, our analysis of the temperature-dependent behavior of sound attenuation in the $J$-$K$-$\Gamma$  model provides a better understanding of the effect of the applied  in-plane field on the sound attenuation
 and helps
better comprehend experimental observations in  $\alpha\text{-}\text{RuCl}_3$~\cite{Hauspurg2024}.
For completeness and comparison, we also present the temperature evolution for the 
for the $K\text{-}\kappa\text{-}\mathbf{h}_{111}$ model in Fig.~\ref{Fig:alpha_temp_evolution}
in App.~\ref{app:alpha-temperature}.



\subsection{Hall viscosity}

Next, we study the phonon dynamics  in the absence of TRS through the Hall viscosity, which probes the Berry curvature in the phonon system~\cite{Avron1995,Barkeshli2012}. Phonons do not couple directly to the magnetic field, but acquire Hall viscosity via spin-lattice coupling, which in the case of the fractionalized Kitaev spin liquid occurs via Majorana fermion-phonon coupling. This interaction introduces a Berry curvature term that mixes transverse and longitudinal phonon modes~\cite{Mengxing2020}, consequently  contributing to the thermal Hall effect~\cite{Shi2012,Aviv2018,Ye2018}.

The phonon Hall viscosity is introduced through the anti-symmetric part of the phonon action in
Eq.~\eqref{eq:antisymmetric_action}. It is a non-dissipative term and represents the leading order effect that breaks time-reversal symmetry.
 Symmetry constraints of the $D_{3d}$ point group result in a single independent Hall viscosity coefficient, denoted by $\eta_H$ \cite{Mengxing2020}. Using linear response theory, $\eta_H$ can be related to the phonon polarization bubble as \cite{Mengxing2020}
\begin{equation}\label{eq:hall_viscosity}
    \eta_H=\frac{1}{q^2\Omega l_a^d}\text{Im}\Pi^{\parallel\perp}(\mathbf{q},\Omega)|_{\Omega\rightarrow 0}.
\end{equation}
Since Hall viscosity results from non-dissipative processes, it requires contributions from off-shell processes in the evaluation of $\Pi^{\parallel\perp}(\mathbf{q},\Omega)$ in Eq.~\eqref{eq:hall_viscosity}. This involves summing contributions from all points in the BZ without enforcing the energy conservation constraint.

\begin{figure}
	\centering
	\includegraphics[width=1.\columnwidth]{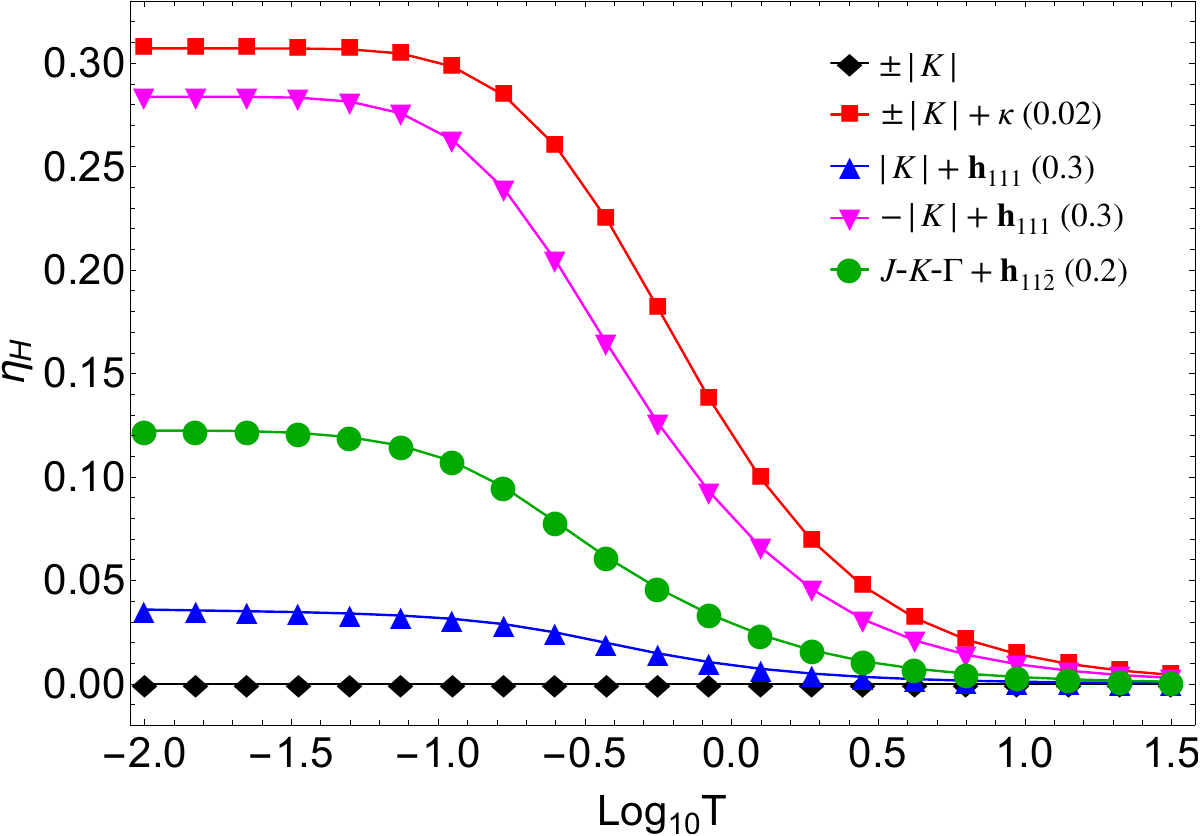}
     \caption{The  temperature dependence of Hall viscosity coefficient $\eta_H$  computed  in different models (see the legend). The model parameter for the green curve are, $K=-1, J=-0.05, \Gamma=0.2$. We set $\bm{q}=(0,0.1\pi)$ and $\Omega=0.2$ with the momentum space discretization of $120\times120$. The artificial energy broadening is $\delta=0.2$. All energies are in the units of $|K|$.
}
\label{Fig:eta_temp_evolution}
\end{figure}

In Fig.\ref{Fig:eta_temp_evolution}, we show the temperature evolution of Hall viscosity for the Kitaev model and the $J$-$K$-$\Gamma$ model with both the $\kappa$ and Zeeman terms. We use a very small sound velocity, $v_s = 0.64$, ensuring $v_s \ll v_F$ for all curves. In this regime, the energy constraints are predominantly satisfied by the ph-processes, with off-shell processes dominated by the pp-processes (as in the pure Kitaev model \cite{Mengxing2020}). In all considered cases, the Hall viscosity shows a similar qualitative behavior: it is maximum at very low (zero) temperature and gradually decreases with increasing temperature. This occurs because the fermionic population of states increases with temperature, gradually suppressing the pp-processes due to the Pauli exclusion principle. Consequently, the Hall viscosity coefficient diminishes at higher temperatures. The differences between the models at low temperatures can be attributed to variations in their corresponding Majorana fermionic spectra.

\section{Conclusions}

 Motivated by recent experiments  in  Kitaev magnets, particularly involving the compound $\alpha$-$\text{RuCl}_3$ under an external magnetic field,  in this work we explored  the phonon dynamics 
 in the extended Kitaev model with an external magnetic field, while remaining in the spin liquid regime.
This extension 
includes the 
non-Kitaev
interactions $J$ and $\Gamma$, and TRS breaking terms, such as the perturbative $\kappa$ interaction and the Zeeman term ${\bf h}$. Since the extended model is not exactly solvable,  in our analysis we have used 
 constrained  self-consistent 
mean-field theory and studied how these terms
  affect the Majorana fermionic spectrum and the flux bands. 
  
  We specifically computed two measurable quantities related to phonon dynamics: sound attenuation, $\alpha_s$, and Hall viscosity, $\eta_H$. These quantities provide insights into different scattering processes, with sound attenuation arising from dissipative scattering and Hall viscosity from non-dissipative processes requiring TRS breaking. Our analysis showed distinct temperature evolution characteristics for these two quantities, consistent with their differing physical origins.

Our analysis also revealed different effects of the TRS breaking terms $\kappa$ and  
   ${\bf h}$. The ground state of the extended Kitaev model  is  a chiral quantum spin liquid  whether the $\kappa$-term, the Zeeman field, or both are present. However, these terms modify the Majorana fermion spectrum differently, leading to varied responses in the phonon dynamics.
 Specifically, the Zeeman term not only opens a gap in the Majorana spectrum at the Dirac  
$\mathbf{K}$-points but also lifts the three-fold degeneracy of the flat gapped modes corresponding to the gauge fermions, causing them to acquire dispersion. 
As a result, the sound attenuation angular distribution in the BZ shows more features. In contrast, the $\kappa$ term opens a larger gap  at the Dirac points, 
 resulting in a more robust Hall viscosity compared with the one in the Zeeman field.

 Finally, our results  for  the $J$-$K$-$\Gamma$ model in the in-plane field ${\bf h}_{11\bar{2}}$ support the observed linear temperature dependence of sound attenuation for the transverse acoustic phonon mode with $v_s<v_F$, as well as the nearly temperature-independent behavior of sound attenuation for the longitudinal acoustic phonon mode with $v_s>v_F$, both of which were reported in a recent ultrasound experiment on $\alpha$-$\text{RuCl}_3$ \cite{Hauspurg2024}. These findings support the idea that the primary cause of sound attenuation at low temperatures is phonon scattering off the fractionalized excitations in Kitaev candidate materials. This provides a more comprehensive picture of the phonon dynamics and the emergent properties of these intriguing systems.

 \section{Acknowledgements}
The authors thank  Vitor  Dantas, Wen-Han Kao, Swetlana Swarup and Yang Yang for valuable discussions. The work is supported by the U.S. Department of Energy, Office of Science, Basic Energy Sciences under Award No. DE-SC0018056.   We acknowledge the support from NSF DMR-2310318 and  the support of the Minnesota Supercomputing Institute (MSI) at the University of Minnesota. N.B.P. also acknowledges the hospitality and partial support  of the Technical University of Munich – Institute for Advanced Study and  the Alexander von Humboldt Foundation.

\appendix
\appendix
\section{Mean-field decomposition}\label{appx:peter_you_better_get_this_appendix_done_already}

 To effectively capture the spin-liquid regime of   not exactly solvable spin Hamiltonian it is common to apply self-consistent parton mean-field theory \cite{Wen2002,Burnell2011,Schaffer2012, Knolle2018, Ferrari2017, Rodrigo2018}. We fractionalize spin degrees of freedom of $\mathcal{H}_s$ into Majorana fermions as indicated in Eq.~(\ref{eq:spin_to_maj}). In this Appendix, we provide the technical details of our approach.


We determine the mean-field ground state by introducing mean-field parameters, such as on-site sublattice magnetizations, which signal the appearance of magnetic order, and bond fields, which characterize the spin liquid state. The resulting mean-field Hamiltonian, given by Eq.~(\ref{eq:MF_spin_ham}), consists solely of Majorana fermion bilinears. This Hamiltonian can be diagonalized numerically in momentum space.
The Fourier transform for Majorana fermions reads
\begin{equation}
    c_{i}=\dfrac{1}{\sqrt{N}}\sum\limits_{\mathbf{k}}\mathrm{e}^{-i \mathbf{k}\cdot\mathbf{R}_i}c_{L,\mathbf{k}},
\end{equation}
where  $\mathbf{R}_i$ denotes the unit cell position of site $i\in$ subblatice $L=A,B$, and the operators  $c_{L,\mathbf{k}}$ satisfy the following algebra:
\begin{equation}
    \{c_{L,\mathbf{k}},c_{L',\mathbf{k}'}\}= 2\delta_{L,L'}\delta_{\mathbf{k},-\mathbf{k}'}, \ \ c_{L,\mathbf{k}}^{\dagger} = c_{L,-\mathbf{k}}.
\end{equation}
In momentum space, the sublattice magnetization and  nearest-neighbor bond mean-fields  from Eq.~(\ref{eq:mfv_in_rspace}) take the form:
\begin{equation}\label{eq:mfv_in_kspace}
    \begin{array}{l}
        \mathrm{m}^{\alpha}_L \!=\! \dfrac{1}{2N}\displaystyle\sum_{\bm{k}}\dfrac{1}{2}\left(\langle ic^{o}_{L,\mathbf{k}}c^{\alpha}_{L,-\mathbf{k}}\rangle +h.c.\right), \\[13pt]
        \mathrm{\Phi}^{a b}_{\delta}\!=\!\dfrac{1}{N}\displaystyle\sum_{\bm{k}}\dfrac{1}{2}\left(\langle ic^{a}_{A,-\mathbf{k}}c^{b}_{B,\mathbf{k}}\mathrm{e}^{-i \mathbf{k}\cdot\bm{\delta}}\rangle+h.c.\right), \\[12pt]
    \end{array}
\end{equation}
where $\bm{\delta}=\bm{R}_j-\bm{R}_i$ is 
the lattice vector difference corresponding to X-, Y-, or Z- bond.
Along with these fields, we also have a set of constrains per site  defined in Eq.~(\ref{eq:constrains}). 
In the mean-field approximation, these constraints are satisfied on average for each sublattice $L=A,B$, 
\begin{equation}\label{eq:avg_con_in_rspace}
    G_{L}^{\alpha} = \frac{1}{2}\sum_{i\in L}\left( i c^{\alpha}_{i} c^o_{i} + \frac{1}{2} \varepsilon^{\alpha\beta\gamma}  i c^{\beta}_{i} c^{\gamma}_{i}\right),  
\end{equation}
where the overall factor of 1/2 does not change the constraints but is introduced for notational convenience later on in the Hamiltonian matrix. To achieve this, we introduced Lagrange multipliers, as detailed in \cite{Ralko2020}, and promoted the mean-field Hamiltonian to:
\begin{equation}\label{eq:mf_and_contrains}
    \mathcal{H}_s \rightarrow \mathcal{H}_s + \sum_{L}\lambda^{\alpha}_L G_{L}^{\alpha}.
\end{equation}
There are six Lagrange multipliers $\lambda^{\alpha}_L$, which have to be determined self-consistently along with the mean fields. The constrains  in the momentum space read
\begin{equation}\label{eq:avg_con_in_kspace}
    G_{L}^{\alpha} =\frac{1}{2} \displaystyle\sum_{\bm{k}}\dfrac{i}{2}\left[ \left( c^{\alpha}_{L,-\mathbf{k}} c^o_{L,\mathbf{k}} + \frac{1}{2} \varepsilon^{\alpha\beta\gamma} i c^{\beta}_{L,-\mathbf{k}} c^{\gamma}_{L,\mathbf{k}}\right) +h.c. \right] \\
\end{equation}
The mean-field Hamiltonian (\ref{eq:MF_spin_ham}) in the momentum space  is block-diagonal. The $8\times8$ matrix, denoted as $H^{\text{MF}}_\mathbf{k}$,
explicitly depends on the mean-field parameters and the Lagrange multipliers and  has the following form:
\begin{equation}\label{eq:MF_kblock_ham}
\begin{array}{c}
H^{\text{MF}}_{\bm{k}} = \left(\begin{array}{cc}
M^{AA} & M^{AB} \\
(M^{AB})^{\dagger} & M^{BB}
\end{array}
\right),
\end{array}
\end{equation}
The explicit form of the $4\times 4$ matices $M^{AA}$, $M^{BB}$,  and $M^{AB}$ in the Majoarana fermion  basis  $\mathcal{C}_{\bm{k}}$ can be easily obtained.
 For example,  $M^{AA}$ is given by
\begin{equation}
M^{AA}=
i\left(\begin{array}{rrrr}
     f^{A}_{o} & -f^{A}_{x} &  -f^{A}_{y} &  -f^{A}_{z} \\
     f^{A}_{x} & \multicolumn{1}{c}{0} &  -g^{A}_{z} &  g^{A}_{y} \\
     f^{A}_{y} & g^{A}_{z} & \multicolumn{1}{c}{0} &  -g^{A}_{x} \\
     f^{A}_{z} & -g^{A}_{y} &  g^{A}_{x} & \multicolumn{1}{c}{0} \\
\end{array}\right)
\end{equation}
with the respective functions
\begin{equation}
\begin{split}
f^{A}_{o}=&i \kappa\sum\limits_{\alpha}\sin(\bm{k}\cdot \bm{\delta}_{A_{\alpha}})\Phi^{\beta\beta}_{\beta}\Phi^{\gamma\gamma}_{\gamma}, \\
f^{A}_{\alpha}=& (3J + K) \mathrm{m}_{B}^{\alpha} + \Gamma (\mathrm{m}_{B}^{\beta}+\mathrm{m}_{B}^{\gamma})+\frac{-h_\alpha + \lambda_{A}^{\alpha}}{4}, \\
g^{A}_{\alpha}=&\frac{-h_\alpha - \lambda_{A}^{\alpha}}{4} ,\\
\end{split}
\end{equation}
with the collection of indices $\alpha[\beta,\gamma]$ follow from cyclic permutations of $x[y,z]$, fixed by the value of $\alpha$. For shortness of notations, we  omit the explicit dependence of matrix elements on $\bm{k}$. 
The matrix elements of $M^{BB}$ can be obtained from those of $M^{AA}$ by replacing sublattice index $A\leftrightarrow B$. 
Lastly, the matrix $M^{AB}$ has the form
\begin{equation}
M^{AB}=
\dfrac{i}{2}\left(\begin{array}{rrrr}
    -f^{AB}_{o}  &  f^{AB}_{ox} &  f^{AB}_{oy} &  f^{AB}_{oz} \\
     f^{AB}_{xo} &  -d^{AB}_{x} &  -g^{AB}_{z} &  -g^{AB}_{y} \\
     f^{AB}_{yo} &  -g^{AB}_{z} &  -d^{AB}_{y} &  -g^{AB}_{x} \\
     f^{AB}_{zo} &  -g^{AB}_{y} &  -g^{AB}_{x} &  -d^{AB}_{z} \\
\end{array}\right)
\end{equation}
with the matrix elements defined as
\begin{equation}
\begin{split}
f^{AB}_{o}=& J\sum\limits_{\alpha}\mathrm{e}^{i\bm{k}\cdot \bm{\delta}_{\alpha}}(\Phi^{xx}_{\alpha} \!+\! \Phi^{yy}_{\alpha} \!+\! \Phi^{zz}_{\alpha}) \!+\! K\sum\limits_{\alpha}\mathrm{e}^{i\bm{k}\cdot \bm{\delta}_{\alpha}}\Phi^{\alpha\alpha}_{\alpha}  \\
& + \Gamma\sum\limits_{\alpha}\mathrm{e}^{i\bm{k}\cdot \bm{\delta}_{\alpha}}(\Phi^{\beta\gamma}_{\alpha} \!+\! \Phi^{\gamma\beta}_{\alpha}),  \\
f^{AB}_{o\alpha}=& (J\!+\!K)\mathrm{e}^{i\bm{k}\cdot\bm{\delta}_\alpha}\mathrm{\Phi}_{\alpha}^{\alpha o}\!+\!J(\mathrm{e}^{i\bm{k}\cdot\bm{\delta}_\beta}\mathrm{\Phi}_{\beta}^{\alpha o}\!+\!\mathrm{e}^{i\bm{k}\cdot\bm{\delta}_\gamma}\mathrm{\Phi}_{\gamma}^{\alpha o})\\
                  & +\Gamma(\mathrm{e}^{i\bm{k}\cdot\bm{\delta}_\beta}\mathrm{\Phi}_{\beta}^{\gamma o}\!+\!\mathrm{e}^{i\bm{k}\cdot\bm{\delta}_\gamma}\mathrm{\Phi}_{\gamma}^{\beta o}), \\
f^{AB}_{\alpha o}=& (J\!+\!K)\mathrm{e}^{i\bm{k}\cdot\bm{\delta}_\alpha}\mathrm{\Phi}_{\alpha}^{o\alpha }\!+\!J(\mathrm{e}^{i\bm{k}\cdot\bm{\delta}_\beta}\mathrm{\Phi}_{\beta}^{o\alpha}\!+\!\mathrm{e}^{i\bm{k}\cdot\bm{\delta}_\gamma}\mathrm{\Phi}_{\gamma}^{o\alpha})\\
                  & +\Gamma(\mathrm{e}^{i\bm{k}\cdot\bm{\delta}_\beta}\mathrm{\Phi}_{\beta}^{o\gamma }\!+\!\mathrm{e}^{i\bm{k}\cdot\bm{\delta}_\gamma}\mathrm{\Phi}_{\gamma}^{o\beta }), \\
g^{AB}_{\alpha}=& \Gamma \mathrm{e}^{i\bm{k}\cdot\bm{\delta}_\alpha}\mathrm{\Phi}_{\alpha}^{oo} ,\\
d^{AB}_{\alpha}=& (J\!+\!K)\mathrm{e}^{i\bm{k}\cdot\bm{\delta}_\alpha}\mathrm{\Phi}_{\alpha}^{oo}\!+\!J(\mathrm{e}^{i\bm{k}\cdot\bm{\delta}_\beta}\mathrm{\Phi}_{\beta}^{oo}\!+\!\mathrm{e}^{i\bm{k}\cdot\bm{\delta}_\gamma}\mathrm{\Phi}_{\gamma}^{oo}),\\
\end{split}
\end{equation}
where the $\bm{\delta}$ has been replaced with $\bm{\delta}_{\alpha}$ for the $\alpha$-bond.

 Next step involves solving the mean-field Hamiltonian through a self-consistent approach. To do this, we  diagonalize  $H^{\text{MF}}_{\bm{k}}$ 
 and use the resulting eigestates $\mathcal{B}_{\bm{k}}$ to evaluate the mean fields from Eq.~(\ref{eq:mfv_in_kspace}).  While unitary diagonalization guarantees the preservation of complex fermion commutation relations, it does not provide the same assurance for Majorana fermions. This  motivates  us to shift to the complex fermion basis.
 
 In real space, the transformation from Majorana fermions to complex fermions can be achieved through the following transformation:
 \begin{equation}\label{eq:fermion_to_majo}
f_{\uparrow}=\frac{1}{2}\left( c^{o} - i c^{z}\right),\, 
f_{\downarrow}=\frac{1}{2}\left( c^{y} - i c^{x}\right), 
\end{equation}
 with $f$-fermions satisfying the usual anticommutation relations: 
$\{f_{s},f_{s'}\}=\{f^{\dagger}_{s},f^{\dagger}_{s'}\}=0$ and $\{f_{s}^{\dagger},f_{s'}\}=\delta_{s,s'}$.
Then, in the momentum space  the change of the basis from the Majorana fermion to complex fermions can be written as
\begin{equation}
\renewcommand{\arraystretch}{0.7}
\begin{array}{c}
    \mathcal{C}_{\bm{k}} = \Sigma \mathcal{F}_{\bm{k}}, \ \  \mathcal{F}_{\bm{k}}\!=\! \left[\begin{array}{c}f_{s,L,\bm{k}} \\  f^{\dagger}_{s,L,-\bm{k}} \end{array}\right]\!,\ \\[0.6cm]
    \Sigma = \left(\begin{array}{rrrrrrrr}
         1 & 0 & 0 & 0 &  1 &  0 &  0 &  0 \\
         0 & i & 0 & 0 &  0 & -i &  0 &  0 \\
         0 & 1 & 0 & 0 &  0 &  1 &  0 &  0 \\
         i & 0 & 0 & 0 & -i &  0 &  0 &  0 \\
         0 & 0 & 1 & 0 &  0 &  0 &  1 &  0 \\
         0 & 0 & 0 & i &  0 &  0 &  0 & -i \\
         0 & 0 & 0 & 1 &  0 &  0 &  0 &  1 \\
         0 & 0 & i & 0 &  0 &  0 & -i &  0 \\
    \end{array}\right),
\end{array}
\end{equation}
where the transformation (pairing) matrix $\Sigma$ is 
 momentum independent. Note that  $\Sigma^{\dagger}\Sigma=2$. 
 Then the diagonalization procedure can be schematically written as
\begin{equation}
\begin{array}{c}
    \mathcal{C}_{-\bm{k}}^{T} H^{\text{MF}}_{\bm{k}} \mathcal{C}_{\bm{k}} = \mathcal{F}_{\bm{k}}^{\dagger} \left( \Sigma^{\dagger} H^{\text{MF}}_{\bm{k}} \Sigma \right) \mathcal{F}_{\bm{k}} \\[12pt]
    \xrightarrow[\text{eigenvectors } U_{\bm{k}} \text{ in ascending order of eigenvalues } E_{\bm{k}}]{\text{Diagonalize in complex fermion basis}} \\[12pt]
    \mathcal{F}_{\bm{k}}^{\dagger} U_{\bm{k}} \left( U_{\bm{k}}^{\dagger} \Sigma^{\dagger} H^{\text{MF}}_{\bm{k}} \Sigma U_{\bm{k}}  \right) U_{\bm{k}}^{\dagger} \mathcal{F}_{\bm{k}} =  \mathcal{B}_{\bm{k}}^{\dagger} E_{\bm{k}} \mathcal{B}_{\bm{k}}.
\end{array}
\end{equation}
The original Majorana basis in momentum space can then be related to the Bogoliubov eigenstates  $\mathcal{B}_{\bm{k}}$
as follows:
\begin{equation}\label{eq:diagonalizng_transformations}
\begin{array}{c}
    \mathcal{B}_{\bm{k}} =  U_{\bm{k}}^{\dagger} \mathcal{F}_{\bm{k}} \Rightarrow  \\
    \mathcal{C}_{\bm{k}} = V_{\bm{k}} \mathcal{B}_{\bm{k}},\ V_{\bm{k}}=\Sigma U_{\bm{k}}.
\end{array}
\end{equation}
 Any expectation values of Majorana bilinears can now be evaluated as:
\begin{equation}\label{eq:diagonalizng_transformations1}
    \langle \mathcal{C}^{i}_{\bm{k}}\mathcal{C}^{j}_{\bm{-k}} \rangle = V_{\bm{k}}^{il}(V_{\bm{k}}^{jl})^*.
\end{equation}
Using this framework, we carry out the mean-field evaluation in 
Eq.~(\ref{eq:mfv_in_kspace}) and verify the average constraints  in Eq.~(\ref{eq:avg_con_in_kspace}).

The complex fermion representation also clarifies the constraints given in the main text in Eq.~(\ref{eq:constrains}), since now we can enumerate states by their occupation number  $n_f = f_{\uparrow}^{\dagger}f_{\uparrow} + f_{\downarrow}^{\dagger}f_{\downarrow}$.
Representing spin by the complex fermions, we have enlarged 
the Hilbert space from a two-dimensional spin space of ``up'' and ``down'' states to a   four-dimensional fermion space of vacuum with $n_f=0$, single occupancy (up and down) with $n_f=1$, and double occupancy $n_f=2$. To address the Hilbert space enlargement, we  enforce single occupancy,  $n_f=1$, which
 redundantly implies that within the physical subspace $f_{\uparrow}^{\dagger}f_{\downarrow}^{\dagger} = f_{\uparrow}f_{\downarrow} = 0$. Expanding these relations into the original Majorana operators results in the three constraints in Eq.~(\ref{eq:constrains}).



\subsection{Mean field self-consistent algorithim}
 To implement the self-consistent  procedure, we employ an iterative approach, following the outlined recipe:
\begin{equation*}
    \begin{array}{@{}rl}
        \multicolumn{2}{l}{\text{\textbf{Main}}}\\
        1.  &    \text{Initial guess }\mathrm{m}, \mathrm{\Phi}, \lambda \\
        2.  &    \delta\mathrm{m}= 1,\delta\mathrm{\Phi} = 1 \\
        3.  &    \text{\textbf{for} } i=1,2,...,\text{MAX} \text{ or } \delta\mathrm{m},\delta\mathrm{\Phi}\!<\!10^{-14}   \\
        4.  &    \ \ \ \ \ \mathrm{m}',\Phi', \langle G \rangle= \text{\textbf{Eval}}(\mathrm{m}, \mathrm{\Phi}, \lambda)   \\
        5.  &    \ \ \ \ \ \text{ \textbf{if} } \mathrm{abs}(\langle G \rangle)\!>\!10^{-14} \\
        6. &    \ \ \ \ \ \ \ \ \ \ \text{Keep } \mathrm{m}, \mathrm{\Phi} \text{ fixed, solve } \langle G(\lambda) \rangle = 0 \text{ for $\lambda$ } \\
        7. &    \ \ \ \ \ \ \ \ \ \ \mathrm{m}',\Phi', \langle G \rangle= \text{\textbf{Eval}}(\mathrm{m}, \mathrm{\Phi}, \lambda)\\
        8. &    \ \ \ \ \ \delta \mathrm{m} = \mathrm{abs}(\mathrm{m}'-\mathrm{m}),\ \delta \Phi = \mathrm{abs}(\Phi'-\Phi) \\
        9. &    \ \ \ \ \ \mathrm{m} = \mathrm{m}',\ \Phi = \Phi' \\
        \multicolumn{2}{l}{\text{\textbf{end}}} \\
    \end{array}
\hspace{1000pt minus 1fill}
\end{equation*}
which makes repeated use of the auxilary \textbf{Eval} function 
\begin{equation*}
\begin{array}{@{}rl}
\multicolumn{2}{l}{\text{\textbf{Eval}}(\mathrm{m}, \mathrm{\Phi}, \lambda)} \\
1.  &    \mathrm{m}', \mathrm{\Phi}', \langle G \rangle=0 \\
2.  &    \text{\textbf{for} every } \bm{k} \\
3.  &    \ \ \ \ \ \text{Diagonalize } H^{\text{MF}}_{\bm{k}}(\mathrm{m},\mathrm{\Phi}, \lambda)\text{ by Eq.~(\ref{eq:diagonalizng_transformations}), and  } \\
    &    \ \ \ \ \ \text{sort the columns of } V_{\bm{k}} \text{ in ascending order} \\
    &    \ \ \ \ \ \text{of eigenvalues} \\
4.  &    \ \ \ \ \ \mathrm{m}' = \mathrm{m}' + .../2N ,\ \Phi' = \Phi' + .../N \text{ (Eq.(\ref{eq:mfv_in_kspace}))} \\
5.  &    \ \ \ \ \ \langle G \rangle = \langle G \rangle + ... /N \text{ (Eq.(\ref{eq:avg_con_in_kspace}))} \\
6.  &    \text{\textbf{return} }\mathrm{m}', \mathrm{\Phi}', \mathrm{G} \\
\multicolumn{2}{l}{\text{\textbf{end}}} \\
\end{array}
\hspace{1000pt minus 1fill}
\end{equation*}
In order to evaluate the mean fields, we discretized the momentum sum with
$\bm{k}= (n_{1}/N_{1})\bm{b}_1 +  (n_{2}/N_{2})\bm{b}_2$, where $\bm{b}_{1(2)}$ are the reciprocal lattice vectors, and $n_{1(2)}\in[0,1,...,N_{1(2)}\!-\!1]$.
The total number of $\bm{k}$ points is thus  $N=N_1N_2$.  We choose rather dense grid  with $N_1=N_2=101$
to avoid special symmetry related $\bm{k}$-points,  which may be singular and need  more careful treatment.
We perform the  numerical diagonalization of $H^{\text{MF}}_{\bm{k}}$ in the momentum space Majorana fermion basis, and multiply fermionic eigenstates by $\sqrt{2}$ to compensate for  
$\Sigma^{\dagger}\Sigma=2$.
The columns of the rescaled diagonalizing matrix are the correct eigenstates, from where we can evaluated all the necessary $\langle \mathcal{C}^{i}_{\bm{k}}\mathcal{C}^{j}_{\bm{-k}} \rangle$ expectation values.

During the evaluation of mean-field variables, the constraints are concurrently examined. 
In the step 5 of Main, if any constraints are found to be violated, we solve for the Lagrange multipliers. This ensures a return to the physical subspace where the constraints are satisfied.
To this end, in  the step 6  of Main, we keep the mean-field variables $\mathrm{m}, \mathrm{\Phi}$ fixed and allow the Lagrange multipliers $\lambda$ to vary such that we solve the equations $\langle G(\lambda) \rangle = 0$. 
There are six equations hidden in this relation, one for every Lagrange multiplier,
and the evaluation depends on the diagonalizing matrix of  $H^{\text{MF}}_{\bm{k}}(\mathrm{m},\mathrm{\Phi}, \lambda)$. Hence, with a given set of Lagrange multipliers, a single evaluation of $\langle G(\lambda) \rangle$ involves calling the Eval function all over again.
Solving for the constraints is by far the most computationally expensive part of the algorithm. We cast the problem as a non-linear minimization, and make use of MINPACK in Fortran, or Ceres library in C++, to minimize the norm of the six equations $||\langle G \rangle||$.
After successfully finding  values  of $\lambda$  that bring us back to the physical subspace, we  use them in  the step 7 of Main and call Eval subroutine  one more time to obtain new values  of $\mathrm{m}'$ and $\mathrm{\Phi}'$.


\section{Details of the magnetoelsic coupling}\label{appx:magnetoelastic_details}

As discussed in Sec.~\ref{sec:magneto-elast}, we use symmetry considerations to construct the magneto-elastic coupling Hamiltonian. Specifically, we employed the $D_{3d}$ point group symmetry to derive Eq.~\eqref{eq:hc_d3d} and the $C_{2h}$ point group symmetry to derive Eq.~\eqref{eq:hc_c2h}. Furthermore, we assumed that the magneto-elastic coupling strength is independent of the symmetry channel and set it equal to $\lambda$. 
 

To obtain the explicit form of the magneto-elastic coupling vertices, we first use the Majorana fermion representation for the spin bilinears $\Xi^{\text{ir}}_{\bm{r},\mathcal{J}}$ in the corresponding  Eq.~\eqref{eq:hc_d3d}  or Eq.~\eqref{eq:hc_c2h}.
We then perform the mean-field decoupling as shown in 
Eq.~\eqref{eq:mf_decouple}. After performing a Fourier transform, for example, for the $D_{3d}$ point group symmetry, we arrive at:
\begin{equation}\label{eq:hc_a1g}
\mathcal{H}_{\text{c}}^{A_{1g}}=\frac{\lambda}{2}\sum_{\textbf{q},\textbf{k}}(iq_au_{a,\textbf{q}}+iq_bu_{b,\textbf{q}})
\bigl[{\mathcal C}_{-\textbf{k}-\textbf{q}}^TH_{\textbf{q},\textbf{k}}^{A_{1g}}{\mathcal C}_{\textbf{k}}\bigr],
\end{equation}
\begin{equation}\label{eq:hc_eg}
\begin{split}
\mathcal{H}_{\text{c}}^{E_g}=&\frac{\lambda}{2}\sum_{\textbf{q},\textbf{k}}\bigl((iq_au_{a,\textbf{q}}-iq_bu_{b,\textbf{q}})\bigl[{\mathcal C}_{-\textbf{k}-\textbf{q}}^TH_{\textbf{q},\textbf{k}}^{E_{g}^1}{\mathcal C}_{\textbf{k}}\bigr]\\
&+(iq_au_{b,\textbf{q}}+iq_bu_{a,\textbf{q}})\bigl[{\mathcal C}_{-\textbf{k}-\textbf{q}}^T H_{\textbf{q},\textbf{k}}^{E_{g}^2}{\mathcal C}_{\textbf{k}}\bigr]\bigr).
\end{split}
\end{equation}
This form captures the symmetry-allowed coupling between the Majorana fermions and the lattice degrees of freedom.
Generically, using the matrix form, we can write $H_{\textbf{q},\textbf{k}}^{\text{ir}}$ as
\begin{equation}
    H_{\text{c}}^{\text{ir}}(\textbf{q},\textbf{k})=\left(\begin{array}{cc}
M_{AA}^{\text{ir}}&M_{AB}^{\text{ir}}\\
M_{BA}^{\text{ir}}&M_{BB}^{\text{ir}}
\end{array}\right),
\end{equation}
where ``ir" labels  the irreducible representation $A_{1g}$, $E_g^1$ or $E_g^2$. The detailed structure of the $4\times 4$ submatrices $M_{AA}^{\text{ir}}, M_{BB}^{\text{ir}}, M_{AB}^{\text{ir}}$  will be given in  the following sections of this Appendix, \ref{app:m_aa_m_bb_IRR} and \ref{app:m_ab_IRR}. 

It is also convenient  for us to express the phonon modes in Eqs.~\eqref{eq:hc_a1g} and \eqref{eq:hc_eg} in terms of the longitudinal and transverse phonon eigenmodes. Then the corresponding coupling Hamiltonian  can be written as
\begin{equation}
\begin{split}
\mathcal{H}_{\textbf{q},\textbf{k}}^{\parallel}=&\tilde{\textbf{u}}_{\textbf{q},\parallel}{\mathcal C}_{-\textbf{k}-\textbf{q}}^T\hat{\lambda}_{\textbf{q},\textbf{k}}^{\parallel}{\mathcal C}_{\textbf{k}},\\
\mathcal{H}_{\textbf{q},\textbf{k}}^{\perp}=&\tilde{\textbf{u}}_{\textbf{q},\perp}{\mathcal C}_{-\textbf{k}-\textbf{q}}^T\hat{\lambda}_{\textbf{q},\textbf{k}}^{\perp}{\mathcal C}_{\textbf{k}},
\end{split}
\end{equation}
where the spin-phonon coupling vertices are
\begin{equation}\label{eq:lambda_lon}
\begin{split}
\hat{\lambda}_{\textbf{q},\textbf{k}}^{\parallel}=&i\frac{\lambda}{2}q H_{\textbf{q},\textbf{k}}^{A_{1g}}
+\frac{\lambda}{2}q\bigl[\cos2\theta_q H_{\textbf{q},\textbf{k}}^{E_{g}^1}+\sin2\theta_q H_{\textbf{q},\textbf{k}}^{E_{g}^2}\bigr],
\end{split}
\end{equation}
\begin{equation}\label{eq:lambda_tr}
\hat{\lambda}_{\textbf{q},\textbf{k}}^{\perp}=i\frac{\lambda}{2}q\bigl[-\sin2\theta_q H_{\textbf{q},\textbf{k}}^{E_{g}^1}+\cos2\theta_q H_{\textbf{q},\textbf{k}}^{E_{g}^2}\bigr].
\end{equation}
We can similarly derive the spin-phonon coupling vertices for the case of a Zeeman field along $\hat{a}$ starting with the spin-phonon coupling Hamiltonian in the irreps of the $C_{2h}$ point group [Eq.\eqref{eq:hc_c2h}-\eqref{eq:hc_c2h_bg}]. However, note that the resulting equations for the coupling vertices will be the same as those given in
 Eqs.~\eqref{eq:lambda_lon} and \eqref{eq:lambda_tr} since there is only a redistribution of the terms among the irreps in the coupling Hamiltonian for the $C_{2h}$ point group compared to the $D_{3d}$ point group.\\

To calculate the phonon polarization bubble, we express the coupling Hamiltonian in the basis of the Bogoliubov quasiparticles using the diagonalizing transformation from
Eq.~\eqref{eq:diagonalizng_transformations}.
The coupling Hamiltonian takes the form:
\begin{equation}
\begin{array}{cc}
   \mathcal{H}_{\textbf{q},\textbf{k}}^{\mu}=\tilde{u}_{\textbf{q},\mu}{\mathcal B}^{\dagger}_{\textbf{k}+\textbf{q}}\tilde{\lambda}_{\textbf{q},\textbf{k}}^{\mu}{\mathcal B}_{\textbf{k}},  \\[0.2cm]
    \tilde{\lambda}_{\textbf{q},\textbf{k}}^{\mu}=V_{\textbf{k}+\textbf{q}}^{\dagger}\lambda_{\textbf{q},\textbf{k}}^{\mu}V_{\textbf{k}}. 
\end{array}
\end{equation}
where $\mu=\parallel,\ \perp$,  and the coupling vertices are divided into four blocks according to the creation and annihilation sectors:
\begin{equation}
\tilde{\lambda}_{\textbf{q},\textbf{k}}^{\mu}=\left[\begin{array}{cc}
\tilde{\lambda}_{\textbf{q},\textbf{k},11}^{\mu}&\tilde{\lambda}_{\textbf{q},\textbf{k},12}^{\mu}\\
\tilde{\lambda}_{\textbf{q},\textbf{k},21}^{\mu}&\tilde{\lambda}_{\textbf{q},\textbf{k},22}^{\mu}
\end{array}\right].
\end{equation}
This formulation allows us to properly account for the interactions between phonons and Bogoliubov quasiparticles, facilitating the calculation of the phonon polarization bubble.

\subsection{Diagonal blocks $M_{AA}^{\text{ir}}$ and $M_{BB}^{\text{ir}}$}\label{app:m_aa_m_bb_IRR}
The  diagonal  block 
 $M_{AA}^{\text{ir}}$  is given by 
\begin{equation}\label{eq:hc_maa}
 M_{AA}^{\text{ir}}= i\left(\begin{array}{cccc}
  0&-f^{A,\text{ir}}_{x}&-f^{A,\text{ir}}_{y}&-f^{A,\text{ir}}_{z}\\
  f^{A,\text{ir}}_{x}&0&0&0\\
    f^{A,\text{ir}}_{y}&0&0&0\\
      f^{A,\text{ir}}_{z}&0&0&0\\
   \end{array}\right).
\end{equation}
 The entries  of $M_{AA}^{\text{ir}}$ have  contributions only from the on-site magnetization decoupling. In the  $A_{1g}$ irrep, they are given by:
 \begin{equation}
    f^{A,A_{1g}}_{\alpha}=2((K+3J)m_B^{\alpha}+\Gamma\sum_{\beta}m_B^{\beta}),
\end{equation}
and for the $E_{g}$ irrep, they are:
\begin{equation}
\begin{split}
    f^{A,E_{g}^1}_{x}=&2(Km_B^{x}+\Gamma(-2m_B^y+m_B^z)),\\
    f^{A,E_{g}^1}_{y}=&2(Km_B^{y}+\Gamma(-2m_B^x+m_B^z)),\\
    f^{A,E_{g}^1}_{z}=&2(-2Km_B^{x}+\Gamma(m_B^x+m_B^y)),\\[0.2cm]
    f^{A,E_{g}^2}_{x}=&2\sqrt{3}(Km_B^{x}-\Gamma m_B^z),\\
    f^{A,E_{g}^2}_{y}=&2\sqrt{3}(-Km_B^{y}+\Gamma m_B^z),\\
    f^{A,E_{g}^2}_{z}=&2\sqrt{3}(\Gamma(m_B^y-m_B^x)).
\end{split}
\end{equation}
The block $M_{BB}^{\text{ir}}$ has the same structure as Eq.~\eqref{eq:hc_maa} with $f^{A,\text{ir}}_{\alpha}\rightarrow f^{B,\text{ir}}_{\alpha}$. The matrix $M^{BB,\text{ir}}$ entries for the $A_{1g}$ irrep are
\begin{equation}
\begin{split}
    f^{B,A_{1g}}_x=&2((K+3J)m_A^{x}e^{i\bm{q}\cdot\bm{n}_1}+\Gamma(m_A^ye^{i\bm{q}\cdot\bm{n}_2}+m_A^z)),\\
    f^{B,A_{1g}}_y=&2((K+3J)m_A^{y}e^{i\bm{q}\cdot\bm{n}_2}+\Gamma(m_A^xe^{i\bm{q}\cdot\bm{n}_1}+m_A^z)),\\  
    f^{B,A_{1g}}_z=&2((K+3J)m_A^{z}+\Gamma(m_A^xe^{i\bm{q}\cdot\bm{n}_1}+m_A^ye^{i\bm{q}\cdot\bm{n}_2})),\\
\end{split}
\end{equation}
where the $\bm{\delta}$ has been replaced with $\bm{\delta}_{x}=\bm{n}_1$ for the X-bond, and similarly $\bm{\delta}_{y}=\bm{n}_2$ and $\bm{\delta}_{z}=\bm{0}$ for Y- and Z-bond respectively, as follows from our unit cell setup in Fig.~\ref{Fig:basic_honey_figure}. The corresponding entries for the $E_{g}$ irrep are
\begin{equation}
\begin{split}
    f^{B,E_{g}^1}_x=&2((J+K)m_A^{x}e^{i\bm{q}\cdot\bm{n}_1}+Jm_A^x(e^{i\bm{q}\cdot\bm{n}_2}-2)\\
    &+\Gamma(-2m_A^y+m_A^ze^{i\bm{q}\cdot\bm{n}_2})),\\
    f^{B,E_{g}^1}_y=&2((J+K)m_A^{y}e^{i\bm{q}\cdot\bm{n}_2}+Jm_A^y(e^{i\bm{q}\cdot\bm{n}_1}-2)\\
    &+\Gamma(m_A^ze^{i\bm{q}\cdot\bm{n}_1}-2m_A^x)),\\
    f^{B,E_{g}^1}_z=&2(-2(J+K)m_A^{z}+Jm_A^z(e^{i\bm{q}\cdot\bm{n}_1}+e^{i\bm{q}\cdot\bm{n}_2})\\
    &+\Gamma(m_A^xe^{i\bm{q}\cdot\bm{n}_2}+m_A^ye^{i\bm{q}\cdot\bm{n}_1})),\\[0.2cm]
    f^{B,E_{g}^2}_x=&2\sqrt{3}((J+K)m_A^{x}e^{i\bm{q}\cdot\bm{n}_1}-Jm_A^{x}e^{i\bm{q}\cdot\bm{n}_2}\\
    &-\Gamma m_B^zm_A^{z}e^{i\bm{q}\cdot\bm{n}_2}),\\
    f^{B,E_{g}^2}_y=&2\sqrt{3}(-(J+K)m_A^{y}e^{i\bm{q}\cdot\bm{n}_2}+Jm_A^{y}e^{i\bm{q}\cdot\bm{n}_1}\\
    &+\Gamma m_B^zm_A^{z}e^{i\bm{q}\cdot\bm{n}_1}),\\
    f^{B,E_{g}^2}_z=&2\sqrt{3}(Jm_A^{z}(e^{i\bm{q}\cdot\bm{n}_1}-e^{i\bm{q}\cdot\bm{n}_2})+\Gamma(m_A^ye^{i\bm{q}\cdot\bm{n}_1}\\
    &-m_A^xe^{i\bm{q}\cdot\bm{n}_2})).    
\end{split}
\end{equation}

\subsection{Off-diagonal block $M^{AB,\text{ir}}$}\label{app:m_ab_IRR}
 The off-diagonal block $M_{AB}^{\text{ir}}$ has contributions from the bond fields and is structured as follows:
\begin{equation}
    M_{AB}^{\text{ir}}=i\left(\begin{array}{rrrr}
    -f^{AB,\text{ir}}_{o}  &  f^{AB,\text{ir}}_{ox} &  f^{AB,\text{ir}}_{oy} &  f^{AB,\text{ir}}_{oz} \\
     f^{AB,\text{ir}}_{xo} &  -d^{AB,\text{ir}}_{x} &  -g^{AB,\text{ir}}_{z} &  -g^{AB,\text{ir}}_{y} \\
     f^{AB,\text{ir}}_{yo} &  -g^{AB,\text{ir}}_{z} &  -d^{AB,\text{ir}}_{y} &  -g^{AB,\text{ir}}_{x} \\
     f^{AB,\text{ir}}_{zo} &  -g^{AB,\text{ir}}_{y} &  -g^{AB,\text{ir}}_{x} &  -d^{AB,\text{ir}}_{z} \\
\end{array}\right).
\end{equation}
For the $A_{1g}$ irrep,
the first matrix entry is:
\begin{equation}
\begin{split}
    f^{AB,A_{1g}}_{o}=&\sum_{\alpha}\bigl( (J+K)\mathrm{\Phi}_{\alpha}^{\alpha\alpha}e^{-i\bm{k}\cdot \bm{\delta}_{\alpha}}+J\sum_{\beta}\mathrm{\Phi}_{\alpha}^{\beta\beta}e^{-i\bm{k}\cdot \bm{\delta}_{\alpha}}\\
    &+\Gamma \sum_{\beta,\gamma}\mathrm{\Phi}_{\alpha}^{\beta\gamma}e^{-i\bm{k}\cdot \bm{\delta}_{\alpha}}\bigr).
\end{split}
\end{equation}
The diagonal entries are:
\begin{equation}
\begin{split}
    d^{AB,A_{1g}}_{\alpha}=&\bigl( (J+K)\mathrm{\Phi}_{\alpha}^{oo}e^{-i\bm{k}\cdot \bm{\delta}_{\alpha}}+J\sum_{\beta}\mathrm{\Phi}_{\beta}^{oo}e^{-i\bm{k}\cdot \bm{\delta}_{\beta}} \bigr).
\end{split}
\end{equation}
The off-diagonal entries are:
\begin{equation}
\begin{split}
    g^{AB,\text{ir}}_{\alpha}=&\Gamma\, \mathrm{\Phi}_{\alpha}^{oo}e^{-i\bm{k}\cdot \bm{\delta}_{\alpha}}.
\end{split}
\end{equation}
The remaining $3 \times 1$ column (and row) matrix entries are:
\begin{equation}
\begin{split}
    f^{AB,A_{1g}}_{o\alpha}=&(J+K)\mathrm{\Phi}_{\alpha}^{\alpha o}e^{-i\bm{k}\cdot \bm{\delta}_{\alpha}}+J\sum_{\beta}\mathrm{\Phi}_{\beta}^{\alpha o}e^{-i\bm{k}\cdot \bm{\delta}_{\beta}}\\
    &+\Gamma\sum_{\beta,\gamma}\mathrm{\Phi}_{\beta}^{\gamma o} e^{-i\bm{k}\cdot \bm{\delta}_{\beta}},\\
     f^{AB,A_{1g}}_{\alpha o}=&(J+K)\mathrm{\Phi}_{\alpha}^{ o\alpha}e^{-i\bm{k}\cdot \bm{\delta}_{\alpha}}+J\sum_{\beta}\mathrm{\Phi}_{\beta}^{o\alpha}e^{-i\bm{k}\cdot \bm{\delta}_{\beta}}\\
    &+\Gamma\sum_{\beta,\gamma}\mathrm{\Phi}_{\beta}^{o\gamma} e^{-i\bm{k}\cdot \bm{\delta}_{\beta}}.
\end{split}
\end{equation}

The first matrix entry for the two components of the $E_g$ irrep are
\begin{equation}
\begin{split}
    f_{o}^{AB,E_{g}^1}=&\bigl((J\!+\!K)\bigl(\mathrm{\Phi}_x^{xx}e^{-i\bm{k}\cdot\bm{n}_1}\!+\!\mathrm{\Phi}_y^{yy}e^{-i\bm{k}\cdot\bm{n}_2}\!-\!2\mathrm{\Phi}_z^{zz}\bigr)\\
    &\!+\!J\bigl((\mathrm{\Phi}_x^{yy}\!+\!\mathrm{\Phi}_x^{zz})e^{-i\bm{k}\cdot\bm{n}_1}\!+\!(\mathrm{\Phi}_y^{xx}\!+\!\mathrm{\Phi}_y^{zz})e^{-i\bm{k}\cdot\bm{n}_2}\\
    &-2(\mathrm{\Phi}_z^{xx}+\mathrm{\Phi}_z^{yy})\bigr)+\Gamma\bigl((\mathrm{\Phi}_x^{yz}+\mathrm{\Phi}_x^{zy})e^{-i\bm{k}\cdot\bm{n}_1}\\
    &+(\mathrm{\Phi}_y^{xz}+\mathrm{\Phi}_y^{zx})e^{-i\bm{k}\cdot\bm{n}_2}-2(\mathrm{\Phi}_z^{xy}+\mathrm{\Phi}_z^{yx})\bigr)\bigr),\\
    f_{o}^{AB,E_{g}^2}=&\sqrt{3}\bigl((J+K)\bigl(\mathrm{\Phi}_x^{xx}e^{-i\bm{k}\cdot\bm{n}_1}-\mathrm{\Phi}_y^{yy}e^{-i\bm{k}\cdot\bm{n}_2}\bigr)\\
    &+J\bigl((\mathrm{\Phi}_x^{yy}+\mathrm{\Phi}_x^{zz})e^{-i\bm{k}\cdot\bm{n}_1} -(\mathrm{\Phi}_y^{xx}+\mathrm{\Phi}_y^{zz})e^{-i\bm{k}\cdot\bm{n}_2} \bigr)\\
    &+\Gamma\bigl((\mathrm{\Phi}_x^{yz}+\mathrm{\Phi}_x^{zy})e^{-i\bm{k}\cdot\bm{n}_1}-(\mathrm{\Phi}_y^{xz}+\mathrm{\Phi}_y^{zx})e^{-i\bm{k}\cdot\bm{n}_2} \bigr)\bigr).
    \end{split}
\end{equation}
The diagonal entries are:
\begin{equation}
    \begin{split}
        d_{x}^{AB,E_{g}^1}=&(J\!+\!K)\mathrm{\Phi}_x^{oo}e^{-i\bm{k}\cdot\bm{n}_1}\!+\!J(\mathrm{\Phi}_y^{oo}e^{-i\bm{k}\cdot\bm{n}_2}\!-\!2\mathrm{\Phi}_z^{oo}),\\
        d_{y}^{AB,E_{g}^1}=&(J\!+\!K)\mathrm{\Phi}_y^{oo}e^{-i\bm{k}\cdot\bm{n}_2}\!+\!J(\mathrm{\Phi}_x^{oo}e^{-i\bm{k}\cdot\bm{n}_1}\!-\!2\mathrm{\Phi}_z^{oo}),\\  
        d_{z}^{AB,E_{g}^1}=&-2(J+K)\mathrm{\Phi}_z^{oo}+J(\mathrm{\Phi}_x^{oo}e^{-i\bm{k}\cdot\bm{n}_1}+\mathrm{\Phi}_y^{oo}e^{-i\bm{k}\cdot\bm{n}_2}),\\[0.2cm]  
        d_{x}^{AB,E_{g}^2}=&\sqrt{3}\bigl((J+K)\mathrm{\Phi}_x^{oo}e^{-i\bm{k}\cdot\bm{n}_1} -J\mathrm{\Phi}_y^{oo}e^{-i\bm{k}\cdot\bm{n}_2}\bigr),\\
        d_{y}^{AB,E_{g}^2}=&-\sqrt{3}\bigl((J+K)\mathrm{\Phi}_y^{oo}e^{-i\bm{k}\cdot\bm{n}_2} -J\mathrm{\Phi}_x^{oo}e^{-i\bm{k}\cdot\bm{n}_1}\bigr),\\
        d_{z}^{AB,E_{g}^2}=&\sqrt{3}J(\mathrm{\Phi}_x^{oo}e^{-i\bm{k}\cdot\bm{n}_1}-\mathrm{\Phi}_y^{oo}e^{-i\bm{k}\cdot\bm{n}_2}).
    \end{split}
\end{equation}
The off-diagonal entries are:
\begin{equation}
    \begin{split}
        &g_x^{AB,E_{g}^1}=\Gamma\mathrm{\Phi}_x^{oo}e^{-i\bm{k}\cdot\bm{n}_1},\\
        &g_y^{AB,E_{g}^1}=\Gamma\mathrm{\Phi}_y^{oo}e^{-i\bm{k}\cdot\bm{n}_2},\\
        &g_z^{AB,E_{g}^1}=-2\Gamma\mathrm{\Phi}_z^{oo},\\[0.2cm]
        &g_x^{AB,E_{g}^2}=\sqrt{3}\Gamma\mathrm{\Phi}_x^{oo}e^{-i\bm{k}\cdot\bm{n}_1},\\
        &g_y^{AB,E_{g}^2}=-\sqrt{3}\Gamma\mathrm{\Phi}_y^{oo}e^{-i\bm{k}\cdot\bm{n}_2},\\
        &g_z^{AB,E_{g}^2}=0.
    \end{split}
\end{equation}
The remaining $1 \times 3$ row matrix entries are:
\begin{equation}
    \begin{split}
        f_{ox}^{AB,E_{g}^1}=&(J+K)\mathrm{\Phi}_x^{xo}e^{-i\bm{k}\cdot\bm{n}_1}+J(\mathrm{\Phi}_y^{xo}e^{-i\bm{k}\cdot\bm{n}_2}-2\mathrm{\Phi}_z^{xo})\\
        &+\Gamma(\mathrm{\Phi}_y^{zo}e^{-i\bm{k}\cdot\bm{n}_2}-2\mathrm{\Phi}_z^{yo}),\\
        f_{oy}^{AB,E_{g}^1}=&(J+K)\mathrm{\Phi}_y^{yo}e^{-i\bm{k}\cdot\bm{n}_2}+J(\mathrm{\Phi}_x^{yo}e^{-i\bm{k}\cdot\bm{n}_1}-2\mathrm{\Phi}_z^{yo})\\
        &+\Gamma(\mathrm{\Phi}_x^{zo}e^{-i\bm{k}\cdot\bm{n}_1}-2\mathrm{\Phi}_z^{xo}),\\  
        f_{oz}^{AB,E_{g}^1}=&\!-\!2(J+K)\mathrm{\Phi}_z^{zo}+J(\mathrm{\Phi}_x^{zo}e^{-i\bm{k}\cdot\bm{n}_1}+\mathrm{\Phi}_y^{zo}e^{-i\bm{k}\cdot\bm{n}_2})\\
        &+\Gamma(\mathrm{\Phi}_x^{yo}e^{-i\bm{k}\cdot\bm{n}_1}+\mathrm{\Phi}_y^{xo}e^{-i\bm{k}\cdot\bm{n}_2}),\\[0.2cm]    f_{ox}^{AB,E_{g}^2}=&\sqrt{3}\bigl( (J+K)\mathrm{\Phi}_x^{xo}e^{-i\bm{k}\cdot\bm{n}_1}-J\mathrm{\Phi}_y^{xo}e^{-i\bm{k}\cdot\bm{n}_2}\\
        &-\Gamma\mathrm{\Phi}_y^{zo}e^{-i\bm{k}\cdot\bm{n}_2} \bigr),\\
        f_{oy}^{AB,E_{g}^2}=&\sqrt{3}\bigl( -(J+K)\mathrm{\Phi}_y^{yo}e^{-i\bm{k}\cdot\bm{n}_2}+J\mathrm{\Phi}_x^{yo}e^{-i\bm{k}\cdot\bm{n}_1}\\
        &+\Gamma\mathrm{\Phi}_x^{zo}e^{-i\bm{k}\cdot\bm{n}_1} \bigr),\\
        f_{oz}^{AB,E_{g}^2}=&\sqrt{3}\bigl( J(\mathrm{\Phi}_x^{zo}e^{-i\bm{k}\cdot\bm{n}_1}-\mathrm{\Phi}_y^{zo}e^{-i\bm{k}\cdot\bm{n}_2})\\
        &+\Gamma(\mathrm{\Phi}_x^{yo}e^{-i\bm{k}\cdot\bm{n}_1}-\mathrm{\Phi}_y^{xo}e^{-i\bm{k}\cdot\bm{n}_2}) \bigr).
    \end{split}
\end{equation}
And, finally,
\begin{equation}
\begin{split}
        f_{xo}^{AB,E_{g}^1}=&(J+K)\mathrm{\Phi}_x^{ox}e^{-i\bm{k}\cdot\bm{n}_1}+J(\mathrm{\Phi}_y^{ox}e^{-i\bm{k}\cdot\bm{n}_2}-2\mathrm{\Phi}_z^{ox})\\
        &+\Gamma(\mathrm{\Phi}_y^{oz}e^{-i\bm{k}\cdot\bm{n}_2}-2\mathrm{\Phi}_z^{oy}),\\
        f_{yo}^{AB,E_{g}^1}=&(J+K)\mathrm{\Phi}_y^{oy}e^{-i\bm{k}\cdot\bm{n}_2}+J(\mathrm{\Phi}_x^{oy}e^{-i\bm{k}\cdot\bm{n}_1}-2\mathrm{\Phi}_z^{oy})\\
        &+\Gamma(\mathrm{\Phi}_x^{oz}e^{-i\bm{k}\cdot\bm{n}_1}-2\mathrm{\Phi}_z^{ox}),\\  
        f_{zo}^{AB,E_{g}^1}=&\!-\!2(J+K)\mathrm{\Phi}_z^{oz}+J(\mathrm{\Phi}_x^{oz}e^{-i\bm{k}\cdot\bm{n}_1}+\mathrm{\Phi}_y^{oz}e^{-i\bm{k}\cdot\bm{n}_2})\\
        &+\Gamma(\mathrm{\Phi}_x^{oy}e^{-i\bm{k}\cdot\bm{n}_1}+\mathrm{\Phi}_y^{ox}e^{-i\bm{k}\cdot\bm{n}_2}),\\[0.2cm]   f_{xo}^{AB,E_{g}^2}=&\sqrt{3}\bigl( (J+K)\mathrm{\Phi}_x^{ox}e^{-i\bm{k}\cdot\bm{n}_1}-J\mathrm{\Phi}_y^{ox}e^{-i\bm{k}\cdot\bm{n}_2}\\
        &-\Gamma\mathrm{\Phi}_y^{oz}e^{-i\bm{k}\cdot\bm{n}_2} \bigr),\\
        f_{yo}^{AB,E_{g}^2}=&\sqrt{3}\bigl( -(J+K)\mathrm{\Phi}_y^{oy}e^{-i\bm{k}\cdot\bm{n}_2}+J\mathrm{\Phi}_x^{oy}e^{-i\bm{k}\cdot\bm{n}_1}\\
        &+\Gamma\mathrm{\Phi}_x^{oz}e^{-i\bm{k}\cdot\bm{n}_1} \bigr),\\
        f_{zo}^{AB,E_{g}^2}=&\sqrt{3}\bigl( J(\mathrm{\Phi}_x^{oz}e^{-i\bm{k}\cdot\bm{n}_1}-\mathrm{\Phi}_y^{oz}e^{-i\bm{k}\cdot\bm{n}_2})\\
        &+\Gamma(\mathrm{\Phi}_x^{oy}e^{-i\bm{k}\cdot\bm{n}_1}-\mathrm{\Phi}_y^{ox}e^{-i\bm{k}\cdot\bm{n}_2}) \bigr).\\         \end{split}
\end{equation}

The matrix $M^{BA,\text{ir}}$ is the same as $(M^{AB,\text{ir}})^{\dagger}$ after replacing $\bm{k}\rightarrow \bm{k}+\bm{q}$.

\section{The phonon polarization bubble}\label{appx:bubble_calculation}

The corrections to the phonon effective action are calculated through the phonon one-loop self-energy, which is given by
\begin{equation}
\begin{split}
    \Pi^{\mu\nu}(\textbf{q},\tau)=&\langle T_{\tau}(\mathcal{B}^{\dagger}_{\textbf{k}+\textbf{q}}\tilde{\lambda}_{\textbf{q},\textbf{k}}^{\mu}\mathcal{B}_{\textbf{k}})(\tau)(\mathcal{B}^{\dagger}_{\textbf{k}'-\textbf{q}}\tilde{\lambda}_{-\textbf{q},\textbf{k}'}^{\nu}\mathcal{B}_{\textbf{k}'})(0)\rangle.
\end {split}
\end{equation}
Then, using Wick's theorem, we get
\begin{equation}
    \begin{split}
        \Pi^{\mu\nu}(\textbf{q},\tau)&=\langle T_{\tau}\mathcal{B}_{\textbf{k}+\textbf{q}}^{\dagger}(\tau)\mathcal{B}_{\textbf{k}'}(0)\rangle \langle T_{\tau}\mathcal{B}_{\textbf{k}}(\tau)\mathcal{B}_{\textbf{k}'-\textbf{q}}^{\dagger}(0)\rangle \tilde{\lambda}_{\textbf{q},\textbf{k}}^{\mu}\tilde{\lambda}_{-\textbf{q},\textbf{k}'}^{\nu}\\
    &-\langle T_{\tau}\mathcal{B}_{\textbf{k}+\textbf{q}}^{\dagger}(\tau)\mathcal{B}_{\textbf{k}'-\textbf{q}}^{\dagger}(0)\rangle \langle T_{\tau}\mathcal{B}_{\textbf{k}}(\tau)\mathcal{B}_{\textbf{k}'}(0)\rangle \tilde{\lambda}_{\textbf{q},\textbf{k}}^{\mu}\tilde{\lambda}_{-\textbf{q},\textbf{k}'}^{\nu}.
    \end{split}
\end{equation}
 Then, by taking $\textbf{k}'=-\textbf{k}$  and selecting appropriate elements of the coupling vertices to complete the fermionic propagators, and subsequently performing the Fourier transform to Matsubara frequency, we arrive at the expression for the phonon one-loop self-energy:
\begin{equation}\label{pi_omega}
    \begin{split}
        \Pi^{\mu\nu}(\mathbf{q},\Omega)=&\text{Tr}[P^{g\bar{g}}_{lm}(-(\textbf{k}+\textbf{q}),-\textbf{k})[\tilde{\lambda}^{\mu}_{\textbf{q},\textbf{k},11}]_{lm}[\tilde{\lambda}^{\nu}_{-\textbf{q},\textbf{k}+\textbf{q},11}]_{ml}\\
        &+P^{\bar{g}g}_{lm}(\textbf{k}+\textbf{q},\textbf{k})[\tilde{\lambda}^{\mu}_{\textbf{q},\textbf{k},22}]_{lm}[\tilde{\lambda}^{\nu}_{-\textbf{q},\textbf{k}+\textbf{q},22}]_{ml}\\
        &+P^{gg}_{lm}(-(\textbf{k}+\textbf{q}),\textbf{k})[\tilde{\lambda}^{\mu}_{\textbf{q},\textbf{k},12}]_{lm}[\tilde{\lambda}^{\nu}_{-\textbf{q},\textbf{k}+\textbf{q},21}]_{ml}\\
        &+P^{\bar{g}\bar{g}}_{lm}(\textbf{k}+\textbf{q},-\textbf{k})[\tilde{\lambda}^{\mu}_{\textbf{q},\textbf{k},21}]_{lm}[\tilde{\lambda}^{\nu}_{-\textbf{q},\textbf{k}+\textbf{q},12}]_{ml}\\
        &-P^{g\bar{g}}_{lm}(-(\textbf{k}+\textbf{q}),-\textbf{k})[\tilde{\lambda}^{\mu}_{\textbf{q},\textbf{k},11}]_{lm}[\tilde{\lambda}^{\nu}_{-\textbf{q},-\textbf{k},22}]_{lm}\\
        &-P^{\bar{g}g}_{lm}(\textbf{k}+\textbf{q},\textbf{k})[\tilde{\lambda}^{\mu}_{\textbf{q},\textbf{k},22}]_{lm}[\tilde{\lambda}^{\nu}_{-\textbf{q},-\textbf{k},22}]_{lm}],\\
         &-P^{gg}_{lm}(-(\textbf{k}+\textbf{q}),\textbf{k})[\tilde{\lambda}^{\mu}_{\textbf{q},\textbf{k},12}]_{lm}[\tilde{\lambda}^{\nu}_{-\textbf{q},-\textbf{k},21}]_{lm}\\
        &-P^{\bar{g}\bar{g}}_{lm}(\textbf{k}+\textbf{q},-\textbf{k})[\tilde{\lambda}^{\mu}_{\textbf{q},\textbf{k},21}]_{lm}[\tilde{\lambda}^{\nu}_{-\textbf{q},-\textbf{k},12}]_{lm}],\\
    \end{split}
\end{equation}
where  $P_{{lm}}^{\bar{g}g}(\textbf{k}+\textbf{q},\textbf{k})$
and similar terms are convolutions of the Matsubara Green's functions of the free fermions $\beta_i$ from the diagonalized mean-field spin Hamiltonian. Here,  $l,m=1,\ldots,4$ correspond to the fermion flavor. 
The term $\text{Tr}[\ldots]$ indicates a trace operation that sums over both the momentum $\textbf{k}$ and the fermion flavors $l,m$.
The Matsubara Green's functions involved in Eq.~\eqref{pi_omega} are given below. We also explicitly show the intermediate step of the Matsubara summation using the residue method \cite{AltlandBook}:
\begin{equation}\label{matsubara_sum_ggbar}
    \begin{array}{rl}
    P_{{l,m}}^{g\bar{g}}= & T\displaystyle\sum\limits_{i\omega_n}g_l(\textbf{k}+\textbf{q},i\omega_n)\bar{g}_m(\textbf{k},i\Omega-i\omega_n) \\
    = & T\displaystyle\sum\limits_{i\omega_n}\dfrac{1}{i\omega_n-\epsilon_{\textbf{k}+\textbf{q},l}}\dfrac{1}{(i\Omega-i\omega_n)+\epsilon_{\textbf{k},m}} \\
    = &\dfrac{n_F(\epsilon_{\textbf{k}+\textbf{q},l})-n_F(\epsilon_{\textbf{k},m})}{i\Omega-\epsilon_{\textbf{k}+\textbf{q},l}+\epsilon_{\textbf{k},m}},
    \end{array}
\end{equation} 
\begin{equation}\label{matsubara_sum_gbarg}
    \begin{array}{rl}  
    P_{{l,m}}^{\bar{g}g}= & T\displaystyle\sum\limits_{i\omega_n}\bar{g}_l(\textbf{k}+\textbf{q},i\omega_n)g_m(\textbf{k},i\Omega-i\omega_n)\\
    = & T\displaystyle\sum\limits_{i\omega_n}\dfrac{1}{i\omega_n+\epsilon_{\textbf{k}+\textbf{q},l}}\dfrac{1}{(i\Omega-i\omega_n)-\epsilon_{\textbf{k},m}}\\
    = & \dfrac{n_F(-\epsilon_{\textbf{k}+\textbf{q},l})-n_F(-\epsilon_{\textbf{k},m})}{i\Omega+\epsilon_{\textbf{k}+\textbf{q},l}-\epsilon_{\textbf{k},m}},
    \end{array}
\end{equation} 
\begin{equation}\label{matsubara_sum_gg}
    \begin{array}{rl}
    P_{{l,m}}^{gg}= & T\displaystyle\sum\limits_{i\omega_n}g_l(\textbf{k}+\textbf{q},i\omega_n)g_m(\textbf{k},i\Omega-i\omega_n)\\
    = & T\displaystyle\sum\limits_{i\omega_n}\dfrac{1}{i\omega_n-\epsilon_{\textbf{k}+\textbf{q},l}}\dfrac{1}{(i\Omega-i\omega_n)-\epsilon_{\textbf{k},m}}\\
    = & \dfrac{n_F(\epsilon_{\textbf{k}+\textbf{q},l})-n_F(-\epsilon_{\textbf{k},m})}{i\Omega-\epsilon_{\textbf{k}+\textbf{q},l}-\epsilon_{\textbf{k},m}},\\
    \end{array}
\end{equation} 
\begin{equation}\label{matsubara_sum_gbargbar}
    \begin{array}{rl}
    P_{{l,m}}^{\bar{g}\bar{g}}= & T\displaystyle\sum\limits_{i\omega_n}\bar{g}_l(\textbf{k}+\textbf{q},i\omega_n)\bar{g}_m(\textbf{k},i\Omega-i\omega_n)\\
    = & T\displaystyle\sum\limits_{i\omega_n}\dfrac{1}{i\omega_n+\epsilon_{\textbf{k}+\textbf{q},l}}\dfrac{1}{(i\Omega-i\omega_n)+\epsilon_{\textbf{k},m}}\\
    = & \dfrac{n_F(-\epsilon_{\textbf{k}+\textbf{q},l})-n_F(\epsilon_{\textbf{k},m})}{i\Omega+\epsilon_{\textbf{k}+\textbf{q},l}+\epsilon_{\textbf{k},m}}.\\
    \end{array}
\end{equation} 
Here we have used the imaginary time free fermion propagators of the spin Hamiltonian, defined as
\begin{equation}
\begin{split}
    g_i(\textbf{k},i\omega_n)=&-\langle T_{\tau}\beta_{\textbf{k},i}(\tau)\beta_{\textbf{k},i}^{\dagger}(0)\rangle_{\omega_n}=\frac{1}{i\omega_n-\epsilon_{\textbf{k},i}},\\
    \bar{g}_i(\textbf{k},i\omega_n)=&-\langle T_{\tau}\beta_{\textbf{k},i}^{\dagger}(\tau)\beta_{\textbf{k},i}(0)\rangle_{\omega_n}=\frac{1}{i\omega_n+\epsilon_{\textbf{k},i}},
\end{split}    
\end{equation}
and $n_F$ is the Fermi-Dirac distribution function.

\section{Temperature evolution of  $\alpha_s^{\parallel}(\textbf{q})$ in  the Kitaev model with different TRS breaking fields }\label{app:alpha-temperature}

\begin{figure}
	\centering
	\includegraphics[width=0.98\columnwidth]{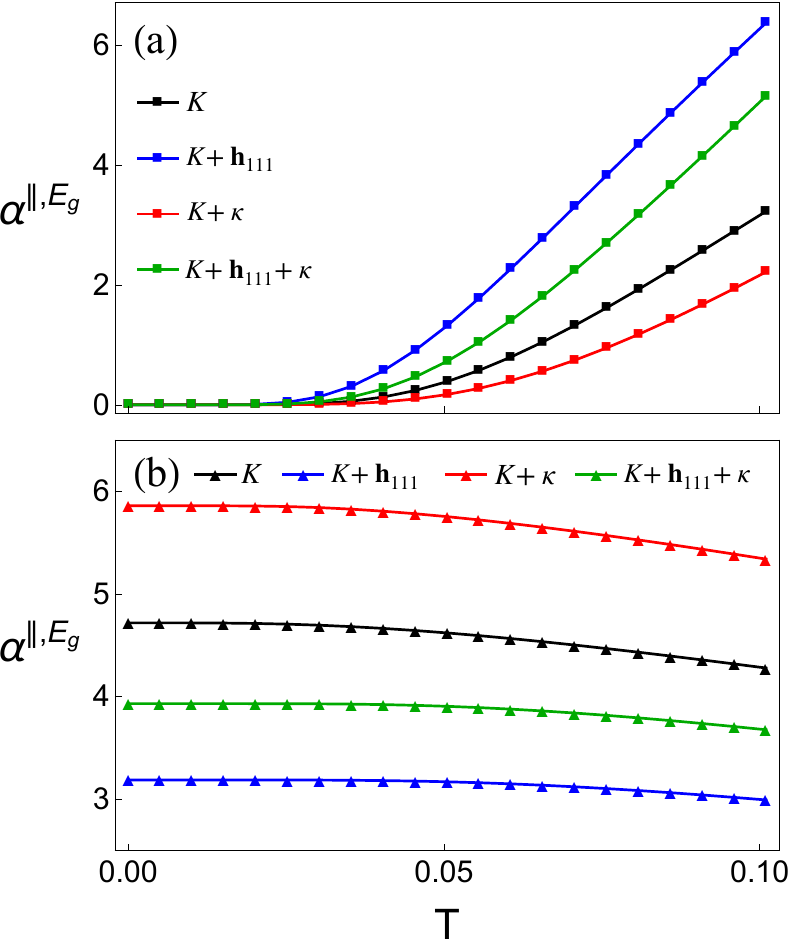}
     \caption{
The temperature evolution of the longitudinal sound attenuation coefficient  $\alpha_s^{\parallel}(\textbf{q})$ for (a) $v_s<v_F$ and (b) $v_s>v_F$ in the $K$, $K+\mathbf{h}_{111}$, $K+\kappa$, and $K+\mathbf{h}_{111}+\kappa$ models is shown. Here, $K=1$, $h_{111}=0.3$, and $\kappa=0.01$. The dominant ph-contribution from the $E_g$ symmetry channel is displayed in (a), and the dominant pp-contribution from the $E_g$ channel is shown in (b). Phonon momenta for (a) and (b) are taken to be $\bm{q}=0.05\pi(\cos\pi/6,\sin\pi/6)$ and $\bm{q}=(0.08\pi,0)$, respectively. For (a), $v_s=0.3$, for (b), $v_s=3.5$, and $v_F=3$. The artificial  energy broadening is set  $\delta=0.2$. All energies are in the units of $|K|$.
}
\label{Fig:alpha_temp_evolution}
\end{figure}

In Fig.~\ref{Fig:alpha_temp_evolution},
we show  temperature evolution of the longitudinal component of the sound attenuation coefficient 
 $\alpha_s^{\parallel}$  in the $K\text{-}\kappa\text{-}\mathbf{h}_{111}$ model for (a) $v_s<v_F$ when the  ph-processes  of attenuations dominate  and (b) $v_s>v_F$  when the  pp-processes prevail. We present results for the pure Kitaev model, $K+\mathbf{h}_{111}$, $K+\kappa$ and $K+\kappa+\mathbf{h}_{111}$ in black, blue, red and green respectively. For the curves in (a), the phonon momentum is fixed to
  $q=0.05\pi$ along $\theta_{\bm{q}}=\pi/6$, a direction where $\alpha_s^{\parallel}$ in the flower shape angular distribution reaches its maximum for 
   $v_s<v_F$ (see Fig~\ref{Fig:alon_khkappa_grid} (a)-(f)). 
   We observe a linear dependence on temperature  for all four cases, albeit with different slopes. 
The relative value of the slopes can be understood from the MF fermionic spectrum in Fig~\ref{Fig:spectrum} (a)-(c). 
In (b), the phonon momentum is fixed to $q=0.08\pi$ along $\theta_{\bm{q}}=0$. Here we see a slowly decreasing sound attenuation with temperature for all four cases. This is because the dominant contribution comes from the pp-processes in this regime.

In summary, the temperature evolution of $\alpha_s^{\parallel}$  for different models in Fig.~\ref{Fig:alpha_temp_evolution} highlights the intricate interplay between the  $\kappa$ term and the $\mathbf{h}_{111}$ Zeeman term, and how these factors modify the Majorana fermion spectra, leading to distinct phonon-fermion interaction dynamics. While these differences affect the overall strength of the sound attenuation, they do not alter the characteristic temperature dependence, which still arises from the fermionic statistics of the Majorana excitations in the spin Hamiltonian.

\bibliography{jour_name_abbreviation.bib,refs.bib}
\end{document}